\documentclass[10pt]{article}

\usepackage[dvips]{graphicx}
\usepackage{amsmath}
\usepackage{amssymb}

\DeclareSymbolFont{AMSb}{U}{msb}{m}{n}
\DeclareMathSymbol{\N}{\mathbin}{AMSb}{"4E}
\DeclareMathSymbol{\Z}{\mathbin}{AMSb}{"5A}
\DeclareMathSymbol{\R}{\mathbin}{AMSb}{"52}
\DeclareMathSymbol{\Q}{\mathbin}{AMSb}{"51}
\DeclareMathSymbol{\I}{\mathbin}{AMSb}{"49}
\DeclareMathSymbol{\C}{\mathbin}{AMSb}{"43}

\begin{document}

\begin{center}

\vspace{0.5cm}
\textbf{\Large Direct demonstration of the completeness of the eigenstates of the Schr{\"o}dinger equation 
               with local and non-local potentials bearing a Coulomb tail}

\vspace{5mm} {\large N.~Michel}

\vspace{3mm}
\textit{Department of Physics, Graduate School of Science, \\ Kyoto University, Kitashirakawa, Kyoto, 606-8502, (Japan)}
\end{center}

\vspace{5mm}

\hrule \vspace{5mm} \noindent{\Large \bf Abstract}
\vspace*{5mm}

Demonstrating the completeness of wave functions solutions of the radial Schr{\"o}dinger equation is a very difficult task.
Existing proofs, relying on operator theory, are often very abstract and far from intuitive comprehension.
However, it is possible to obtain rigorous proofs amenable to physical insight, if one restricts the considered class of Schr{\"o}dinger potentials.
One can mention in particular unbounded potentials yielding a purely discrete spectrum and short-range potentials.
However, those possessing a Coulomb tail, very important for physical applications, have remained problematic due
to their long-range character. The method proposed in this paper allows to treat them correctly, provided the non-Coulomb part of potentials
vanishes after a finite radius. Non-locality of potentials can also be handled.
The main idea in the proposed demonstration is that regular solutions behave like sine/cosine functions for large momenta, 
so that their expansions verify Fourier transform properties.
The highly singular point at $k = 0$ of long-range potentials is dealt with properly using analytical properties of Coulomb wave functions.
Lebesgue measure theory is avoided, rendering the demonstration clear from a physical point of view.

\section{Introduction}
Completeness of wave functions of the Schr{\"o}dinger equation has always been a difficult subject in quantum mechanics.
Due to the importance of eigenfunctions expansions in quantum physics, a global comprehension of the phenomena underlying the property
of completeness is necessary. 

Demonstrations of completeness have been treated by many authors. Fourier and Fourier-Bessel series come back to the works
of Dirichlet for the former \cite{Dirichlet} and Hankel, Schl{\"a}fli and Young for the latter \cite{Hankel_Schlafli_Young} 
(see Ref.\cite{Watson} for a thorough study of Bessel functions properties and demonstration of completeness).
Both proofs use complex analysis only. Fourier series demonstration is standard and rely on Dirichlet kernel properties.
The Fourier-Bessel series proof is similar to the former through the use of a generalization of the Dirichlet kernel.
The general case of differential equations defined in a finite interval, of which Fourier and Fourier-Bessel expansions are a particular case,
is treated within regular Sturm-Liouville theory \cite{Birkhoff,Titchmarsh}.
Completeness can therein be demonstrated using Rayleigh's quotient \cite{Morse_Feschbach,Kemble} or closeness arguments of the considered
set of eigenstates with another basis \cite{Birkhoff}, with which, however, point convergence cannot be handled.
The regular Sturm-Liouville problem with non-local potentials has been treated by Fubini and Lichtenstein \cite{Fubini_Lichtenstein}.
Extension to the singular case of infinite intervals, albeit for differential equations only,
has first been considered by Weyl \cite{Weyl}, and continued afterward by Titchmarsh \cite{Titchmarsh}.
The study of spectral theorem of the general linear operator can be found in Ref.\cite{Dunford_Schwarz}.
If one restricts operators to be short-range spherical potentials for the radial Schr{\"o}dinger equation, 
the method of Newton \cite{Newton} can be applied.
It deals with complex integration only and, thus, has the advantage to be straightforward mathematically.
Its extension to the Coulomb case was, however, not considered. It has been effected recently, but only for the case of Coulomb wave functions, for which
analytical properties of the confluent hypergeometric function can be employed \cite{Mukhamedzhanov}.

Newton's completeness relation has played a fundamental role in the development of non-hermitian formalism, 
as it is the starting point of the Berggren completeness relation \cite{Berggren}, where bound, resonant and complex scattering states are used.
It has been shown to be a powerful tool to expand strongly correlated nuclear states \cite{PRL_michel,PRC1_michel,PRC2_michel}.
With a continuous spectrum, the theory of rigged Hilbert spaces \cite{R_de_la_Madrid1,R_de_la_Madrid2} 
is necessary to be able to use Dirac formalism correctly. This theory cannot be built unless completeness of eigenstates wave functions has been proved.

Newton's completeness relation extended to the proton case with repulsive Coulomb potential was demonstrated in Ref.\cite{PRC2_michel}, 
using discrete completeness relations given by box boundary conditions, becoming continuous letting the radius of the box go to infinity.
However, mathematical details were missing, which could not be treated in Ref.\cite{PRC2_michel} as the principal motive of the paper
was the study of loosely bound and resonant nuclear systems and not the completeness demonstration by itself. They will be described in this paper.
The aim of this paper is also to extend the demonstration of Ref.\cite{PRC2_michel} to potentials with attractive Coulomb tails.
The main difficulty therein is to treat properly the infinite set of bound states accumulating at zero energy for attractive potentials.
The generalization to the non-hermitian case of complex potentials will be also be performed.
Standard mathematical results of fundamental importance for our demonstration will be presented in appendices.

\section{Hamiltonian potentials} \label{H_pot}
One will analyze the sets of eigenstates provided by one-body Schr{\"o}dinger equation Hamiltonians.
One will call eigenstates both bound and scattering states for simplicity, 
even though the denomination is improper for non-square integrable positive energy states.
We will first consider the radial Schr{\"o}dinger equation for real potentials, 
non-hermitian complex potentials being handled afterward.
Potentials can be non-local, with their local part denoted by $v(r)$ and their non-local part by $w(r,r')$.
They are demanded to vanish identically for $r > R_0$, $r' > R_0$, except for pure Coulombic asymptotic, which can be attractive or repulsive.
$v(r)$ and $w(r,r')$ uphold the following conditions:
\begin{eqnarray}
&&v(r) = \frac{V_c}{r} \mbox{ , } r > R_0, \label{v_asymp_inf} \\
&&w(r,r') \sim w(0,r') r^{\ell+1} \mbox{ , } r \rightarrow 0 \mbox{ and } w(r,r') \rightarrow 0 \mbox{ , } r \rightarrow R_0~~\forall r' \in [0:R_0],
\label{w_zero_cont_cond}
\end{eqnarray}
where $V_c \in \R$. $v(r)$ will be supposed to be continuous except maybe for a finite set of radii, where it has to be integrable, 
and $w(r,r')$ integrable on its domain of definition,
unless explicitly stated. $w(r,r')$ is naturally supposed to be symmetric in $r$ and $r'$.
The first condition in Eq.(\ref{w_zero_cont_cond}) is necessary for eigenstates to be well-defined 
and is automatically fulfilled for Hartree-Fock potentials \cite{Vautherin_Veneroni}.
Demanded conditions are virtually always verified in practice, as potentials subtracted from their Coulomb component
usually go very quickly to zero at large distance.
The studied Schr{\"o}dinger equation then reads:
\begin{eqnarray}
u''(r) = \left( \frac{\ell(\ell+1)}{r^2} + v(r) - k^2 \right) u(r) + \int_{0}^{R_0} w(r,r')~u(r')~dr' \label{Schrodinger_eq}
\end{eqnarray}
where $u(r)$ is the radial wave function, not necessarily eigenstate of the Hamiltonian, defined by $v(r)$ and $w(r,r')$,
$\ell$ is its angular momentum and $k$ is its linear momentum. $\ell$ can bear non-integral values in our demonstration but is restricted
for the moment to $\ell \geq 0$. Negative values of $\ell$, for which one can always consider 
$\ell \geq -\frac{1}{2}$ due to invariance of Eq.(\ref{Schrodinger_eq}) with $\ell \rightarrow -\ell-1$, 
will be effected in the context of non-hermitian potentials.
Existence and unicity of $u(r)$ functions in Eq.(\ref{Schrodinger_eq}) for arbitrary $k$ is provided for $r \geq R_0$ by the Cauchy-Lipshitz theorem, 
as there Eq.(\ref{Schrodinger_eq}) is local and $v(r)$ continuous except at mentioned singularities, whose treatment is standard.
For $r < R_0$, it can be equivalently written as an integral equation which then can be handled by Fredholm theory as $R_0$ is finite
and $v(r)$, $w(r,r')$ are integrable. Hamiltonians bearing an effective mass are implicitly accounted for with Eq.(\ref{Schrodinger_eq})
as they can always be rewritten to verify a Schr{\"o}dinger equation bearing no effective mass via a point-canonical transformation 
\cite{Alhaidari}. The considered effective mass must evidently become constant for $r > R_0$.

The Hamiltonian defined in Eq.(\ref{Schrodinger_eq}) in open radial interval 
possesses a lower bound for eigenstate energy, i.e.~a bound ground state or a continuum 
of positive energy states only. This property is immediate for local potentials, 
with which convexity of wave functions for sufficiently large $-k^2$ values is used to demonstrate this property \cite{Kemble}.
As non-locality in our class of potentials vanishes after $R_0$, it is possible to use this type of argument for non-local potentials.
Indeed, if a wave function, regular in $r = 0$ but not necessarily for $r \rightarrow +\infty$, 
has its last node in $[R_0:+\infty[$, 
making its energy more negative will either send this node to infinity and have it disappeared
or make it converge to a finite radius $R_1 \geq R_0$. 
Hence, there exists in all possible cases a negative energy $e_{min}$ and radius $R_1 > 0$
for which all wave functions sustaining $e < e_{min}$ are nodeless in $[R_1:+\infty[$.
Thus, the Hamiltonian box spectrum for $R > R_1$ has a ground state which will converge either to the Hamiltonian ground state in open radial interval
or become part of the continuum if the open radial interval spectrum has no discrete part therein.

\section{Large momentum expansion of wave functions} \label{large_k}
The fundamental idea of the proposed demonstration is that wave functions possess an analytical expansion in the limit of large momentum.
Formulas are based on the semi-classical approximation inspired from Wentzel-Brillouin-Kramer (WKB) theory \cite{Messiah}. 
In Ref.\cite{PRC2_michel}, however, spherical Bessel functions are used instead of the usual sine/cosine functions.
Their use, indeed, allows to take into account properly the first turning point induced by the centrifugal barrier in Eq.(\ref{Schrodinger_eq}), so that it is
available for all $r \geq 0$, $k$ sufficiently large. One will demonstrate its validity in this section.
The simplest way for this purpose is to introduce the error equation \cite{WKB_error}, i.e.~one writes the exact solution of Eq.(\ref{Schrodinger_eq})
as a product of approximated solution and error term.
We will suppose for the moment that $v(r)$ and $w(r,r')$ are twice differentiable.
In order to study our semi-classical approximation of Eq.(\ref{Schrodinger_eq}), we define the following functions:
\begin{eqnarray}
&&\lambda(r) = \sqrt{k^2 - v(r)} \mbox{ , } \Lambda(r) = \int_{0}^{r} \lambda(r')~dr' \label{lambdas_def_large_k}, \\
&&u_{f_0}(r) = \lambda(r)^{-\frac{1}{2}}~f_{\ell}(\Lambda(r)) \mbox{ , } u_f(r) = u_{f_0}(r)~[1+\epsilon(r)], \label{exact_function}
\end{eqnarray}
where $k$ is the linear momentum of the wave function $u(r)$ of Eq.(\ref{Schrodinger_eq}),
$f_{\ell}$ is a spherical Riccati-Bessel function, denoted as $\hat{j}_{\ell}$, $h^{+}_{\ell}$ or $h^{-}_{\ell}$ 
respectively regular, outgoing and incoming wave solutions of Eq.(\ref{Schrodinger_eq}) for $v(r)$ and $w(r,r')$ identically equal to zero,
$u_{f_0}(r)$ is the approximated solution and $\epsilon(r)$ the error function of Eq.(\ref{Schrodinger_eq}).
$k$ can be arbitrary large (which one will denote also as $k > k_{min}$) so as to have $\lambda(r)$, $\Lambda(r)$ and $u_{f_0}(r)$ defined $\forall r \geq 0$.
For the method to work, one needs $u_{f_0}(r)$ never to vanish except maybe at $r=0$.
Hence, one will have $f_{\ell} = \hat{j}_{\ell}$ if $r \in [0:r_0]$, $r_0$ chosen small enough so that $f_{\ell}(\Lambda(r)) \neq 0$ for $r \in ]0:r_0]$,
and $f_{\ell} = h^{\pm}_{\ell}$ if $r > r_0$.
Inserting Eq.(\ref{exact_function}) in Eq.(\ref{Schrodinger_eq}), one obtains the error equation verified by $\epsilon(r)$ with $r>0$:
\begin{eqnarray}
\epsilon''(r) &+& \epsilon'(r) \left[ \frac{d}{dx} \log(u_{f_0}(x)^2) \right]_{x=r} = F(k,r,\epsilon) \label{error_eq_large_k}, \\
F(k,r,\epsilon) &=& [1 + \epsilon(r)] \left( \frac{1}{2} \lambda''(r) \lambda(r)^{-1} - \frac{3}{4} \lambda'(r)^2 \lambda(r)^{-2} \right) \nonumber \\
                &+& \ell(\ell+1) [1 + \epsilon(r)] \left( \frac{1}{r^2} - \frac{k^2 - v(r)}{\Lambda(r)^2} \right)  \nonumber \\
                &+& \frac{1}{u_{f_0}(r)} \int_{0}^{R_0} w(r,r') u_{f_0}(r') [1 + \epsilon(r')]~dr' \label{F_def}.
\end{eqnarray}
Note that the term depending on $\ell(\ell+1)$ appears in Eq.(\ref{F_def}) because of the use of Bessel functions in Eq.(\ref{exact_function}).
One can solve Eq.(\ref{error_eq_large_k}) as a first-order equation according to $\epsilon'(r)$ treating $F(k,r,\epsilon)$
as formally independent of $\epsilon(r)$. This provides the integral equation verified by $\epsilon(r)$:
\begin{eqnarray}
\epsilon(r) = \int_{r_d}^{r} \frac{dr'}{u_{f_0}(r')^2}
              \int_{r_d}^{r'} u_{f_0}(r'')^2 F(k,r'',\epsilon)~dr'', \label{int_eq}
\end{eqnarray}
where $r_d = 0$ for $f_{\ell} = \hat{j}_{\ell}$ and $r_d = +\infty$ for $f_{\ell} = h^{\pm}_{\ell}$.
The iterative Picard's method to formally solve Eq.(\ref{int_eq}) comes forward:
\begin{eqnarray}
\epsilon_{0}(r) = 0 \mbox{ , } 
\epsilon_{n+1}(r) = \int_{r_d}^{r} \frac{dr'}{u_{f_0}(r')^2} \int_{r_d}^{r'} u_{f_0}(r'')^2 F(k,r'',\epsilon_{n})~dr'' \mbox{ , } n \geq 0, \label{eps_Picard}
\end{eqnarray}
with which $\epsilon'_{n+1}(r)$ is obtained by a simple differentiation. 
One will show that $\epsilon_{n}(r) \rightarrow \epsilon(r)$ for $n \rightarrow +\infty$
and $\epsilon(r) = O(k^{-2})$, $\epsilon'(r) = O(k^{-2})$ when $k \rightarrow +\infty$ by recurrence on $n \in \N$, and that $\forall r \geq 0$.
$\epsilon_{0}(r) = O(k^{-2})$ and $\epsilon'_{0}(r) = O(k^{-2})$ are trivially verified.
One thus supposes $\epsilon_{n}(r) = O(k^{-2})$ and $\epsilon'_{n}(r) = O(k^{-2})$ when $k \rightarrow +\infty$, defined $\forall r \geq 0$, and this
property has to be demonstrated for $\epsilon_{n+1}(r)$ and $\epsilon'_{n+1}(r)$.
$F(k,r,\epsilon) = O(\log(r) r^{-3})$ for $r \rightarrow +\infty$ in Eq.(\ref{F_def}), 
so that $\epsilon_{n+1}(r)$ is defined with $r_d = +\infty$ in Eq.(\ref{eps_Picard}).
The non-local term involving $w(r,r')$ in Eq.(\ref{F_def}) converges in $r' = 0$ and has a finite limit for $r \rightarrow 0$
because of Eq.(\ref{w_zero_cont_cond}), 
$u_{f_0}(r) \sim r^{\ell+1}$ or $u_{f_0}(r) \sim r^{-\ell}$ equivalents (up to one constant)
for $r \rightarrow 0$ and the fact that $\epsilon_{n}(0)$ is finite.
$\epsilon_{n+1}(r)$ in Eq.(\ref{eps_Picard}) remains finite for $r \rightarrow 0$, even though $F(k,r,\epsilon_{n})$ is undefined in $r=0$
by way of the term proportional to $\ell(\ell+1)$, which is $O(r^{-1})$ for $r \rightarrow 0$.
This is demonstrated using Eq.(\ref{w_zero_cont_cond}) and $u_{f_0}(r) \sim r^{\ell+1}$ or $u_{f_0}(r) \sim r^{-\ell}$ equivalents for $r \rightarrow 0$ as well.
The same argument holds for $\epsilon'_{n+1}(r)$.
From Eqs.(\ref{lambdas_def_large_k},\ref{F_def}), one can see that $F(k,r,\epsilon_{n})$ is the sum of two terms.
one being the product of $1 + \epsilon_{n}(r)$ multiplied by a function easily shown to be $O(k^{-2})$ when $k \rightarrow +\infty$, 
and the other involving only the non-local potential $w(r,r')$. Due to finiteness of $\epsilon_{n}(r)$ for $r \geq 0$,
the first term, integrated in Eq.(\ref{eps_Picard}), will remain $O(k^{-2})$ when $k \rightarrow +\infty$.
The other term involving $w(r,r')$ is treated as in Ref.\cite{PRC2_michel}, with two integrations by parts, using Eq.(\ref{w_zero_cont_cond}):
\begin{eqnarray}
\frac{1}{u_{f_0}(r)} \int_{0}^{R_0} w(r,r')~u_{f_0}(r')~dr' 
= - \frac{1}{u_{f_0}(r)} \left[ \frac{\partial w}{\partial r'}(r,r')~\mathcal{U}_{f_0}(r') \right]^{r'=R_0}_{r'=0}
+ \frac{1}{u_{f_0}(r)} \int_{0}^{R_0} \frac{\partial^2 w}{\partial r'^2}(r,r')~\mathcal{U}_{f_0}(r')~dr', \label{eps_non_loc_part_int_1}
\end{eqnarray}
where $\mathcal{U}_{f_0}(r)$ is defined so that $\mathcal{U}_{f_0}''(r) = u_{f_0}(r)$, 
$\{\mathcal{U}_{f_0}(r_0),\mathcal{U}_{f_0}'(r_0)\}$ chosen so as to have $\mathcal{U}_{f_0}(r) u_{f_0}(r)^{-1} = O(k^{-2})$
for $k \rightarrow +\infty$, choice rendered possible via the use of $f_{\ell}$ asymptotic behavior for large arguments \cite{Abramowitz_Stegun}.
As $w(r,r')$ is bounded as twice differentiable on $[0:R_0]^2$, the non-local term of Eq.(\ref{eps_non_loc_part_int_1}) is
$O(k^{-2})$ when $k \rightarrow +\infty$.
The remaining part of the non-local term in Eq.(\ref{eps_Picard}) involving $\epsilon_{n}(r')$ reads after one integration by part:
\begin{eqnarray}
\frac{1}{u_{f_0}(r)} \int_{0}^{R_0} w_{\epsilon_{n}}(r,r')~u_{f_0}(r')~dr' = 
-\frac{1}{u_{f_0}(r)} \int_{0}^{R_0} \frac{\partial w_{\epsilon_{n}}}{\partial r'}(r,r')~\mathcal{U}_{f_0}'(r')~dr', 
\label{eps_non_loc_part_int_2}
\end{eqnarray}
where $w_{\epsilon_{n}}(r,r')$ = $w(r,r')~\epsilon_{n}(r')$.
As $\mathcal{U}_{f_0}'(r) u_{f_0}(r)^{-1} = O(k^{-1})$ and as it is assumed that $\epsilon_{n}(r) = O(k^{-2})$ and $\epsilon'_{n}(r) = O(k^{-2})$
when $k \rightarrow +\infty$, the integral in Eq.(\ref{eps_non_loc_part_int_2}) is $O(k^{-3})$ when $k \rightarrow +\infty$.
Thus, $\epsilon_{n+1}(r) = O(k^{-2})$ and $\epsilon'_{n+1}(r) = O(k^{-2})$ when $k \rightarrow +\infty$.
Using Eqs.(\ref{error_eq_large_k},\ref{F_def},\ref{eps_Picard},\ref{eps_non_loc_part_int_1},\ref{eps_non_loc_part_int_2}), 
the following equation holds using recurrence on $n \in \N$:
\begin{eqnarray}
\epsilon_{n}(r) = \frac{E(r,k)}{k^2} + O \left( \frac{1}{k^3} \right) \mbox{ , } n \in \N \mbox{ , } k \rightarrow +\infty, \label{eps_n_asymp}
\end{eqnarray}
where $E(r,k)$ is independent of $n$ and bounded $\forall k \geq K$, $K$ large enough, by a constant independent of $k$,
while $O(k^{-3})$ is bounded $\forall n \in \N$ by a constant independent of $n$.
Thus, for $k \geq K$, $K$ large enough, $\epsilon(r)$ can always be written by way of a converging Neumann-Liouville power series in $k^{-1}$.
Hence, as all terms in Eq.(\ref{eps_n_asymp}) can be majored by constants independent of $n$, Eq.(\ref{eps_n_asymp}) is valid for $\epsilon(r)$ as well.

Noting the clear linear independence of both $u_{f}(r)$ functions involving $f_{\ell} = h^{\pm}_{\ell}$ for $k$ large enough,
one can now express $u(r)$, eigenstate of Eq.(\ref{Schrodinger_eq}), in terms of the $u_{f}(r)$ functions of Eq.(\ref{exact_function}). 
$u(r)$ has to vanish in $r=0$, so that one can write without loss of generality:
\begin{eqnarray}
u(r) &=& \mathcal{N}~\lambda(r)^{-\frac{1}{2}}~[\hat{j}_{\ell}(\Lambda(r))+\epsilon(r)] \mbox{ , } r < r_0, \label{u_exp_small_r} \\
u(r) &=& \lambda(r)^{-\frac{1}{2}} \left[ \mathcal{N}^+~(h^+_{\ell}(\Lambda(r))+\epsilon^{+}(r))
     + \mathcal{N}^-~(h^-_{\ell}(\Lambda(r))+\epsilon^{-}(r)) \right] \mbox{ , } r \geq r_0, \label{u_exp_large_r}
\end{eqnarray}
where $\mathcal{N}$, $\mathcal{N}^+$, $\mathcal{N}^-$ are normalization constants and $\epsilon(r)$, $\epsilon^{+}(r)$ and $\epsilon^{-}(r)$
the functions defined in Eq.(\ref{exact_function}) for respectively $f_{\ell} = \hat{j}_{\ell}$ in Eq.(\ref{u_exp_small_r})
and $f_{\ell} = h^{\pm}_{\ell}$ in Eq.(\ref{u_exp_large_r}), multiplied by $f_{\ell}$ for convenience.
By continuity of $u(r)$, its expressions in Eq.(\ref{u_exp_small_r}) and Eq.(\ref{u_exp_large_r}) 
must be equal at $r_0 > 0$, which can be chosen to be arbitrarily small.
It readily implies that $\mathcal{N} = 2i \mathcal{N}^+ = - 2i\mathcal{N}^-$ and $2i \epsilon(r) = \epsilon^{+}(r) - \epsilon^{-}(r)$ 
in Eqs.(\ref{u_exp_small_r},\ref{u_exp_large_r}).
Thus, $\epsilon(r) = O(k^{-2})$ $\forall r \geq 0$ when $k \rightarrow +\infty$ in Eq.(\ref{u_exp_small_r}).
Eq.(\ref{u_exp_small_r}) provides the asymptotic expansion in $k^{-1}$ of Ref.\cite{PRC2_michel}:
\begin{eqnarray}
&&u(r) = C_k \hat{j}_{\ell} (kr) - C_k \frac{\mathcal{V}(r)}{2k} \hat{j}'_{\ell} (kr) + O \left( \frac{C_k}{k^2} \right) 
       \mbox{ , } k \rightarrow +\infty, \label{u_large_k} \\
&&\mathcal{V}(r) = \int_{0}^{r} v(r')~dr' \label{v_primitive},
\end{eqnarray}
where $C_k$ is a normalization factor.
$u(r)$ and the two first terms of the right-hand side of Eq.(\ref{u_large_k}) are bounded $\forall r \geq 0$ and $k \rightarrow +\infty$
by a constant depending only on $\mathcal{V}(r)$ and $C_k$.
This property immediately holds for $O \left( \frac{C_k}{k^2} \right)$ in Eq.(\ref{u_large_k}).
Thus, Eq.(\ref{u_large_k}) is still valid when lifting the condition of twice differentiability
for $v(r)$ and $w(r,r')$ for $(r,r') \in [0:R_0]^2$, as the space of twice-differentiable functions is dense with respect to 1-norm
in the set of potentials verifying the conditions stated in Sec.(\ref{H_pot}). 
If $k \rightarrow +\infty$ and $r \rightarrow +\infty$, it is preferrable to use the following expansion, 
deduced from Eq.(\ref{eps_n_asymp}) and Eq.(\ref{u_exp_small_r}) as well:
\begin{eqnarray}
u(r) = C_k \hat{j}_{\ell} (\Lambda(r)) + C_k \frac{\alpha_k(r)}{k^2} \label{u_large_k_r},
\end{eqnarray}
where $\sup_{k > k_{min}} |\alpha_k(r)| \rightarrow 0$ when $r \rightarrow +\infty$.

\section{Cancellation of Dirac delta distributions} \label{compl_rel_diff_box}
As described in Ref.\cite{PRC2_michel}, completeness relations involving continuous set of states are handled with box discrete completeness
relations, whose radius $R$ will go to $+\infty$. Used box boundary conditions are $u(0) = u(R) = 0$, which define uniquely eigenstates wave functions.
From the fact that the Hamiltonian of Eq.(\ref{Schrodinger_eq}) possesses therein a ground state (see Sec.(\ref{H_pot})),
the completeness of a discrete set of eigenstates of Eq.(\ref{Schrodinger_eq}) defined in a finite interval $[0:R]$ 
can be procured for scalar product norm \cite{Morse_Feschbach,Kemble}. 
Pointwise convergence for our class of Hamiltonians will be a consequence of the results obtained in this section.

The completeness relation of eigenstates of the Hamiltonian of Eq.(\ref{Schrodinger_eq}) 
with box boundary conditions in a finite interval $[0:R]$ formally reads:
\begin{eqnarray}
\sum_{n \in b} u_n(r) u_n(r') + \sum_{m=1}^{+\infty} u(k_m,r) u(k_m,r') (k_{m} - k_{m-1}) = \delta(r - r') \label{formal_H_comp_rel}
\end{eqnarray}
where the $u_n(r)$ states are called ``bound'', as they will become square-integrable on the real axis for $R \rightarrow +\infty$,
while the $u(k_m,r)$ states are called ``scattering'', as they will converge to non-integrable scattering states for $R \rightarrow +\infty$.
Bound states of positive energy, called bound states embedded in the continuum (BSEC) \cite{BSEC},
are allowed, but bound states of energy zero are forbidden. This last point is clearly of no consequence for practical calculations.
The number of BSEC's for our class of Hamiltonians will be demonstrated 
to be finite and independent of $R$ as long as $R > R_0$ in Sec.(\ref{large_R_sec}), so that they will pose no problem when $R \rightarrow +\infty$. 
Wave functions are normalized by way of the following equalities:
\begin{eqnarray}
&&\int_{0}^{R} u_n(r)^2~dr = 1 \mbox{ , } \int_{0}^{R} u(k_m,r)^2~dr = \frac{1}{k_{m} - k_{m-1}}, \label{u_n_km_norm}
\end{eqnarray}
where $k_{m} - k_{m-1}$ is used instead of $k_{m+1} - k_{m}$ in Ref.\cite{PRC2_michel} as it is more convenient for attractive Coulomb case.
Intuitively, for $R \rightarrow +\infty$, the sum of Eq.(\ref{formal_H_comp_rel}) involving bound states will remain finite (repulsive/no Coulomb case) 
or become infinite (attractive Coulomb case), while its infinite series built from $u(k_m,r)$ states will become an integral.
One can also foresee the appearance of Dirac delta normalization of scattering states
in Eq.(\ref{u_n_km_norm}). The limiting process cannot be done directly, however, 
as Eq.(\ref{formal_H_comp_rel}) is purely formal due to the divergent character of its series.
In order to avoid convergence problems, a new method has been introduced in Ref.\cite{PRC2_michel}.
It consists in subtracting the completeness relation generated
by Fourier-Bessel series to the one of Eq.(\ref{formal_H_comp_rel}). One then generates a convergent series for fixed $(r,r')$,
so that the transformation from series to integral when $R \rightarrow +\infty$ can be effected properly.

One defines Bessel discretized states and formal completeness relation analogously to Eqs.(\ref{formal_H_comp_rel},\ref{u_n_km_norm}):
\begin{eqnarray}
\sum_{m=1}^{+\infty} B_{\kappa_m}^2 \hat{j}_{\ell}(\kappa_m r) \hat{j}_{\ell}(\kappa_m r') (\kappa_{m} - \kappa_{m-1}) 
                     = \delta(r - r') \label{formal_Bessel_comp_rel} \label{Bessel_comp_rel} \\
B_{\kappa_m}^2 \int_{0}^{R} \hat{j}_{\ell}(\kappa_m r)^2~dr = \frac{1}{\kappa_{m} - \kappa_{m-1}}. \label{Bessel_km_norm}
\end{eqnarray}
where $B_{\kappa_m}$ is a normalization constant and $\kappa_m$, $m \in \N^*$, is a discretized linear momentum of the Fourier-Bessel series, 
verifying $\hat{j}_{\ell}(\kappa_m R) = 0$. One recalls the asymptotic expansion of $\hat{j}_{\ell}(x)$ when $x \rightarrow +\infty$:
\begin{eqnarray}
\hat{j}_{\ell}(x) = \sin \left( x - \frac{\pi}{2} \ell \right) - \frac{a_{\ell}}{2 x} \cos \left( x - \frac{\pi}{2} \ell \right) 
                 + O \left( \frac{1}{x^2} \right) \label{jl_asymp}
\end{eqnarray}
where $a_{\ell} = -\ell(\ell+1)$.
Subtracting Eq.(\ref{formal_Bessel_comp_rel}) from Eq.(\ref{formal_H_comp_rel}) provides:
\begin{eqnarray}
S_R(r,r') &=& \sum_{n \in b} u_n(r) u_n(r') \nonumber \\
&+& \sum_{m=1}^{+\infty} \left[ u(k_m,r) u(k_m,r') (k_{m} - k_{m-1})
- B_{\kappa_m}^2 \hat{j}_{\ell}(\kappa_m r) \hat{j}_{\ell}(\kappa_m r') (\kappa_{m} - \kappa_{m-1}) \right] \label{formal_conv_series}
\end{eqnarray}
where $S_R(r,r')$ is for the moment a distribution only known to be equal to zero in a weak sense.
It is necessary to show that the series of Eq.(\ref{formal_conv_series})
converges $\forall (r,r') \in [0:R]^2$, after which one can infer that $S_R(r,r') = 0$ in a strong sense.
For this, one will rewrite the main results of Ref.\cite{PRC2_michel}. 

We first determine the asymptotic expansion of $k_m$ and $\kappa_m$ when $m \rightarrow +\infty$. 
From Eqs.(\ref{u_large_k},\ref{jl_asymp}), one obtains for $m \rightarrow +\infty$:
\begin{eqnarray}
k_m = \frac{\left( m + \frac{\ell}{2}\right) \pi}{R} + \frac{a_{\ell} + R \mathcal{V}(R)}{2 R m \pi} + O \left( \frac{1}{m^2} \right) \mbox{ , }
\kappa_m = \frac{\left( m + \frac{\ell}{2}\right) \pi}{R} + \frac{a_{\ell}}{2 R m \pi} + O \left( \frac{1}{m^2} \right) \label{km_kappa_m_asymp}
\end{eqnarray}
The condition that the same integer $m$ enters both linear momenta expansions of Eq.(\ref{km_kappa_m_asymp}) without finite shift
is enforced demanding that $u(k_m,r)$ and $\hat{j}_{\ell}(\kappa_m r)$ have both $m$ nodes, 
having for consequence that a finite number of $u(k_m,r)$ in Eq.(\ref{formal_conv_series}) vanish if bound states are present therein.
The $C_{k_m}$ and $B_{\kappa_m}$ constants expansion defined with Eqs.(\ref{u_large_k},\ref{u_n_km_norm},\ref{Bessel_km_norm}) 
are calculated with Eqs.(\ref{u_large_k},\ref{jl_asymp}), using the fact that the integral of Eq.(\ref{Bessel_km_norm}) is analytical:
\begin{eqnarray}
&&- B_{\kappa_m}^2 \hat{j}_{\ell+1}(\kappa_m R) \hat{j}_{\ell-1}(\kappa_m R) = \frac{2}{R (\kappa_{m} - \kappa_{m-1})}, \label{Bm_exact} \\
&&C_{k_m}^2 \left( \left[ \hat{j}_{\ell}(k_m R)^2 - \hat{j}_{\ell+1}(k_m R) \hat{j}_{\ell-1}(k_m R) \right]
- \int_0^{R} \frac{2 \mathcal{V}(r)}{k_m R} \hat{j}_{\ell}(k_m r) \hat{j}'_{\ell}(k_m r)~dr + O \left( \frac{1}{k_m^2} \right) \right)  \nonumber \\
&&= \frac{2}{R (k_{m} - k_{m-1})}. \label{Cm_asymp}
\end{eqnarray}
Eq.(\ref{jl_asymp}) and the fact that the integral involving $\mathcal{V}(r)$ in Eq.(\ref{Cm_asymp}) 
is $O(k_m^{-2})$ when $k_m \rightarrow +\infty$, which can be seen with a partial integration, provide for $m \rightarrow +\infty$:
\begin{eqnarray}
C_{k_m} = \sqrt{\frac{2}{\pi}} + O(m^{-2}) \mbox{ , } B_{\kappa_m} = \sqrt{\frac{2}{\pi}} + O(m^{-2}) \label{Cm_Bm_two_pi}
\end{eqnarray}
The general term of the series of Eq.(\ref{formal_conv_series}) 
reads for $m \rightarrow +\infty$, by way of Eqs.(\ref{u_large_k},\ref{jl_asymp},\ref{km_kappa_m_asymp},\ref{Cm_Bm_two_pi}):
\begin{eqnarray}
&&u(k_m,r) u(k_m,r') (k_{m} - k_{m-1}) - B_{\kappa_m}^2 \hat{j}_{\ell}(\kappa_m r) \hat{j}_{\ell}(\kappa_m r') (\kappa_{m} - \kappa_{m-1}) \nonumber \\
&=& \frac{R(\mathcal{V}(r) - \mathcal{V}(r')) - (r - r') \mathcal{V}(R)}{2 m \pi R}  
\sin \left( \frac{\pi m}{R} (r - r') + \ell \pi \left( \frac{r - r'}{2R} \right) \right) \nonumber \\
&-& \frac{R(\mathcal{V}(r) + \mathcal{V}(r')) - (r + r') \mathcal{V}(R)}{2 m \pi R}  
\sin \left( \frac{\pi m}{R} (r + r') + \ell \pi \left( \frac{r + r'}{2R} - 1 \right) \right)
+ O (m^{-2}). \label{series_term_asymp} 
\end{eqnarray}
The non-absolutely convergent part of Eq.(\ref{series_term_asymp})
consists in a standard Fourier series term. Abel transformation can be performed on Eq.(\ref{series_term_asymp}) 
in order to have it normally convergent with respect to $r$ and $r'$ (see App.(\ref{Abel_transformation})).
As a consequence, $S_R(r,r')$ is finite and continuous $\forall (r,r') \in [0:R]^2$.
Thus, owing to Eqs.(\ref{formal_conv_series},\ref{series_term_asymp}):
\begin{eqnarray}
\int_{0}^{R} S_R(r,r')~u(r')~dr' &=& \sum_{n \in b} u_n(r) \int_{0}^{R} u_n(r') u(r')~dr'
+ \sum_{m=1}^{+\infty} \left[ u(k_m,r)  (k_{m} - k_{m-1}) \int_{0}^{R} u(k_m,r') u(r')~dr'\right. \nonumber \\
&-& \left. B_{\kappa_m}^2 \hat{j}_{\ell}(\kappa_m r) (\kappa_{m} - \kappa_{m-1}) \int_{0}^{R} \hat{j}_{\ell}(\kappa_m r') u(r')~dr' \right]
= u(r) - u(r) = 0 \label{SR_norm},
\end{eqnarray}
where $u(r)$ is a state equal to $u_n(r)$, $n \in \N$ or $u(k_m,r)$, $m \in \N^*$ (see Eq.(\ref{formal_H_comp_rel})).
This result is obtained by way of orthonormality of the set of $u_n(r)$ and $u(k_m,r)$ states, pointwise completeness of the set of Bessel functions
and by inverting series and integral integrating Eq.(\ref{formal_conv_series}) between $0$ and $R$, 
which is justified due to the normal convergence obtained by Abel transformation of the series of Eq.(\ref{formal_conv_series}).
As the set of $u_n(r)$ and $u(k_m,r)$ states is complete for scalar product norm, $\int_{0}^{R} S_R(r,r')^2~dr' = 0$.
As $S_R(r,r')$ is finite and continuous $\forall (r,r') \in [0:R]^2$, $S_R(r,r') = 0$ $\forall (r,r') \in [0:R]^2$.

One will show that it implies pointwise completeness of the set of $u_n(r)$ and $u(k_m,r)$ states in Eq.(\ref{formal_H_comp_rel}).
One considers an arbitrary wave function $f(r)$, assumed to yield a Fourier-Bessel series on $[0:R]$.
Its expansion with the set of $u_n(r)$ and $u(k_m,r)$ states is equal $\forall r \in [0:R]$ to its Fourier-Bessel series expansion,
which can be seen calculating $\int_{0}^{R}S_R(r,r')~f(r') ~dr'$ similarly as in Eq.(\ref{SR_norm}).
The set of $u_n(r)$ and $u(k_m,r)$ functions then possesses the same properties as Fourier-Bessel, 
and hence Fourier series \cite{Watson}, i.e.~the generalized Fourier expansion of $f(r)$ 
provided by Eq.(\ref{formal_H_comp_rel}) is equal to $\lim_{\delta \rightarrow 0} \frac{f(r+\delta) + f(r-\delta)}{2}$.

\section{Wave functions behavior for box radius $R \rightarrow +\infty$} \label{large_R_sec}
In order to be able to let $R$ go to $+\infty$ in Eq.(\ref{formal_conv_series}), one has to show first that wave functions
converge to their open radial interval counterparts. The case of bound states of negative energy 
can be treated considering the appearance of nodes at $r=R$ when $|k^2|$ decreases in Eq.(\ref{Schrodinger_eq}) \cite{Kemble}, 
as Eq.(\ref{Schrodinger_eq}) is local at large distance (see Sec.(\ref{H_pot})). The fact that they are square-integrable on $[0:+\infty[$ 
readily implies uniform convergence of a box bound wave function $u_n(r)$, $n \in \N$ fixed, with respect to $r \in [0:+\infty[$ when $R \rightarrow +\infty$ 
to its limit defined in open radial interval. The number of bound states of negative energy will become infinite only for potentials with attractive Coulomb tail
in our class of Hamiltonians. Indeed, it occurs only if the number of nodes of the regular solution $u(k=0,r)$ is itself infinite \cite{Messiah}.
This can be proved with WKB approximations of $u(k=0,r)$ for $r \rightarrow +\infty$ (see Sec.(\ref{small_k_attractive}) for the Coulomb attractive case).

The case of scattering states is different, 
as one has to show both the accumulation of box scattering states in $]0:+\infty[$ for $R \rightarrow +\infty$ and
the consistency of their normalization by way of Eq.(\ref{u_n_km_norm}). For their study, one considers a linear momentum $k > 0$
and $R$ always chosen larger than $R_0$.
Eq.(\ref{v_asymp_inf}) implies that the wave function $u(k,r)$ of linear momentum $k > 0$ behaves for $r > R_0$ as:
\begin{eqnarray}
u(k,r) = \sqrt{\frac{2}{\pi}} \mathcal{N}_k \left( S^+ h^+_{\ell \eta} (k r) + S^- h^-_{\ell \eta} (k r) \right), \label{u_large_R_h_pm}
\end{eqnarray}
where $\eta$ is the Sommerfeld parameter, here equal to $\frac{V_c}{2k}$, $h^{\pm}_{\ell \eta}$ is the outgoing/incoming irregular
Coulomb wave function \cite{Abramowitz_Stegun} and $\mathcal{N}_k$, $S^+$ and $S^-$ are normalization constants 
chosen so that $\mathcal{N}_k \in \R$ and $S^+ S^- = \frac{1}{4}$
(reality of wave functions imply that $S^+ = (S^-)^*$). The $\sqrt{\frac{2}{\pi}}$ factor is introduced in order to have
$\mathcal{N}_k \rightarrow 1$  for scattering states when $R \rightarrow +\infty$.
Due to Eq.(\ref{u_large_R_h_pm}), wave functions for $k > 0$ are not square integrable in $[0:+\infty[$ unless $\mathcal{N}_k = 0$ therein.
This implies that all BSEC's of the open radial interval problem are present in the box spectrum 
and independent of $R$ once $R > R_0$, as they have to uphold box boundary
conditions for $R = R_0$ and vanish identically afterward. 
Moreover, it was demonstrated in Sec.(\ref{large_k}) that all wave functions of linear momentum $k$ large enough verify 
Eq.(\ref{u_large_k}) for fixed box radius $R$, so that it is impossible for wave functions of arbitrarily large $k$
to vanish identically in $[R_0:R]$ and not in $[0:R_0]$.
Thus, BSEC's have to be in finite number and can be all put in the bound states sum of Eq.(\ref{formal_H_comp_rel}).
We can thus always suppose that $k$ is not a BSEC momentum, implying $\mathcal{N}_k \neq 0$.
Using asymptotic expansion of Coulomb wave functions for large arguments in Eq.(\ref{u_large_R_h_pm}) \cite{Abramowitz_Stegun}, one obtains:
\begin{eqnarray}
u(k,r) &=& \sqrt{\frac{2}{\pi}} \mathcal{N}_k \left[ 
\sin(k r - \eta \log (2 k r) + \delta(k)) + \frac{\ell(\ell+1)}{2kr} \cos(k r - \eta \log (2 k r) + \delta(k)) 
+ O \left( \frac{1}{k^2 r} \right) \right] \label{u_large_R}
\end{eqnarray}
with $\delta(k) \in [0:2 \pi[$ the phase shift associated with $u(k,r)$ and the $k$ dependence is written explicitly in the rest term
to allow $k \rightarrow + \infty$ along with $r \rightarrow + \infty$ at the end of the calculation.
For $u(k_n,r)$ eigenstate of Eq.(\ref{Schrodinger_eq}), $n \in \N^*$, implying $u(k_n,R) = 0$, the following equality holds from Eq.(\ref{u_large_R}):
\begin{eqnarray}
k_n = \frac{n \pi - \delta_n}{R} + \frac{V_c~\log(2 k_n R)}{2 k_n R} - \frac{\ell(\ell+1)}{2 k_n R} + O \left( \frac{1}{k_n^2 R^2} \right), \label{kn_eq_large_R}
\end{eqnarray}
where $\eta_n$ and $\delta_n$ are its respective Sommerfeld parameter and phase shift, nevertheless modified so that one might have $\delta_n \neq \delta(k_n)$.
Indeed, in Eq.(\ref{kn_eq_large_R}), $\delta_n$ has been defined so that Eqs.(\ref{km_kappa_m_asymp},\ref{kn_eq_large_R})
provide the same value of linear momentum for $k_n$ and $k_m$ if one has $n=m$ in both equations.
For $R \rightarrow + \infty$, $k_n \rightarrow k$ and $\delta_n$ converge to a finite value for $n \in \N$ appropriately chosen, 
as $u(k_n,r)$ differs significantly from its asymptotic value on Eq.(\ref{u_large_R}) only in a finite region of the real axis.
One can see from Eq.(\ref{kn_eq_large_R}) that for $R \rightarrow +\infty$:
\begin{eqnarray}
k_{\nu} \rightarrow 0 \mbox{ , } \nu \in \N^* \mbox{ , } \nu = o(R), \label{kn_small_nu_large_R}
\end{eqnarray}
as supposing therein the opposite for $\sup(k_{\nu})_{R' > R}$ leads to a contradiction.
For $R$ large enough, it is thus possible to choose $n$ so that $k_n \leq k < k_{n+1}$.
The difference $k_{n+1} - k_{n}$ reads by way of Eq.(\ref{kn_eq_large_R}):
\begin{eqnarray}
k_{n+1} - k_n &=& \frac{\pi}{R} - \frac{\delta_{n+1} - \delta_n}{R}
                          + \frac{V_c}{2R} \left( \frac{\log(2 k_{n+1} R)}{k_{n+1}} - \frac{\log(2 k_n R)}{k_n} \right)  \nonumber \\
                          &-& \frac{\ell(\ell+1)}{2R} \left( \frac{1}{k_{n+1}} - \frac{1}{k_n} \right)
                          + O \left( \frac{1}{k_n^2 R^2} \right). \label{kn_diff_detailed}
\end{eqnarray}
Hence, $k_{n+1} - k_{n} \rightarrow 0$ in Eq.(\ref{kn_diff_detailed}) for $R \rightarrow +\infty$, 
and consequently $\delta_{n+1} - \delta_{n} \rightarrow 0$ as well.
As $k_{n+1} \rightarrow k$ and $k_n \rightarrow k$ when $R \rightarrow +\infty$, Eq.(\ref{kn_diff_detailed}) can be simplified to:
\begin{eqnarray}
k_{n+1} - k_{n} = \frac{\pi}{R} + o \left( \frac{1}{R} \right), \label{kn_diff}
\end{eqnarray}
This proves that the set of $\{k_n\}_{n \in \N^*}$ becomes uniformly dense in $]0:+\infty[$ when $R \rightarrow +\infty$
(see Eqs.(\ref{kn_small_nu_large_R},\ref{kn_diff_detailed},\ref{kn_diff})).

We can now show that Eq.(\ref{u_n_km_norm}) provides the standard Dirac delta normalization of scattering functions for $R \rightarrow +\infty$.
Inserting Eqs.(\ref{u_large_R},\ref{kn_diff}) in Eq.(\ref{u_n_km_norm}), one can derive for $R \rightarrow +\infty$:
\begin{eqnarray}
\mathcal{N}_{k_n}^2 
\left[ \int_{R_0}^{R} \left[ \sin(k_n r - \eta \log (2 k_n r) + \delta(k_n)) + O \left( \frac{1}{k_n r} \right) \right]^2~dr + O(1) \right]
= \frac{R}{2} + o(R), \label{norm_equiv_large_R}
\end{eqnarray}
where the part of the integral from $0$ to $R_0$ is accounted for by the $O(1)$ term 
of the left-hand side of Eq.(\ref{norm_equiv_large_R}).
The latter integral is bounded when $R \rightarrow +\infty$, and hence $k_n \rightarrow k$, because $u(k_n,0) = 0$ 
and $\mathcal{N}_{k_n}^{-1} u(k_n,R_0)$ has a limit therein from Eq.(\ref{u_large_R_h_pm}), 
implying $\mathcal{N}_k^{-1} u(k,r)$ boundedness in $[0:R_0]$ for $R \rightarrow +\infty$.
The left-hand side integral in Eq.(\ref{norm_equiv_large_R}) is dealt with similarly to Eq.(\ref{Cm_asymp}).
Consequently, we arrive to the standard value of $\mathcal{N}_{k}$ of scattering states for $R \rightarrow +\infty$, equivalent to Dirac delta normalization:
\begin{eqnarray}
\mathcal{N}_{k} = 1 + o(1), \label{norm_large_R}
\end{eqnarray}
uniformly in all intervals $[k_d:+\infty[$, $k_d > 0$, and in $]0:+\infty[$ if $\ell = V_c = 0$ 
as only the principal term of Eq.(\ref{u_large_R}) remains in this case.
As all solutions of Eq.(\ref{Schrodinger_eq}) regular at $r=0$ for fixed $k$ and $R > R_0$ are proportional, one has proved the convergence of box
scattering states to the continuum of positive energy states of Eq.(\ref{Schrodinger_eq}) verifying Dirac delta normalization.

Comparing Eq.(\ref{u_large_k_r}) with $r \rightarrow +\infty$ to Eq.(\ref{u_large_R}) with $k \rightarrow +\infty$, one obtains:
\begin{eqnarray}
C_k = \sqrt{\frac{2}{\pi}} \mathcal{N}_k. \label{Ck_delta_large_R_large_k}
\end{eqnarray}
The expansion of $k_m$ issued from Eq.(\ref{u_large_k_r}) and Eqs.(\ref{kn_diff},\ref{norm_large_R}), deduced from it, read in this regime:
\begin{eqnarray}
k_m &=& \frac{(m + \frac{\ell}{2}) \pi}{R} + \frac{a_{\ell} + R \mathcal{V}(R)}{2 k_m R^2} + \frac{\beta_{k_m}(R)}{R k_m^2}, \nonumber \\
k_{n+1} - k_{n} &=& \frac{\pi}{R} + \frac{Y_k(R)}{k^2 R} \mbox{ , } \mathcal{N}_{k} = 1 + \frac{Z_k(R)}{k^2},
\label{kn_diff_norm_large_R_large_k}
\end{eqnarray}
where $\sup_{k > k_{min}} |\beta_{k}(R)|$ has a finite limit for $R \rightarrow +\infty$ in Eq.(\ref{kn_diff_norm_large_R_large_k}),
$\sup_{k > k_{min}} |Y_k(R)|$ and $\sup_{k > k_{min}} |Z_k(R)|$ vanish for $R \rightarrow +\infty$.
The corresponding $\mathcal{N}_{k}$ expansion is procured from Eqs.(\ref{u_large_k_r}), (\ref{u_n_km_norm}) and (\ref{kn_diff_norm_large_R_large_k}) noticing that:
\begin{eqnarray}
\int_{0}^{R} \hat{j}_{\ell}^2 (\Lambda(r))~dr &=& 
\left[ \hat{j}_{\ell}(\Lambda(R))^2 - \hat{j}_{\ell+1}(\Lambda(R)) \hat{j}_{\ell-1}(\Lambda(R)) \right] \frac{\Lambda(R)}{2k} \nonumber \\
&+& \int_{0}^{R} \hat{j}_{\ell}^2 (\Lambda(r)) \left[ 1 - \frac{\lambda(r)}{k}\right]~dr \nonumber \\
&=& \frac{R}{2} + \frac{R \alpha_{k}(R)}{k^2} \label{jl2_Lambda_integral},
\end{eqnarray}
with $\sup_{k > k_{min}} |\alpha_{k}(R)| \rightarrow 0$ for $R \rightarrow +\infty$.
The Bessel function case, embodied in Eqs.(\ref{km_kappa_m_asymp},\ref{Cm_Bm_two_pi}), 
is an obvious particular case of Eq.(\ref{kn_diff_norm_large_R_large_k}):
\begin{eqnarray}
\kappa_m &=& \frac{(m + \frac{\ell}{2}) \pi}{R} + \frac{a_{\ell}}{2 \kappa_m R^2} + \frac{\beta_{\kappa_m}(R)}{R \kappa_m^2}, \nonumber \\
\kappa_{n+1} - \kappa_{n} &=& \frac{\pi}{R} + \frac{Y_{\kappa}(R)}{\kappa^2 R} \mbox{ , } B_\kappa = \sqrt{\frac{2}{\pi}} + \frac{Z_{\kappa}(R)}{\kappa^2}
\label{Bessel_large_R_large_k}.
\end{eqnarray}
where $\sup_{\kappa > \kappa_{min}} |\beta_\kappa(R)|$ has a finite limit for $R \rightarrow +\infty$, 
and $\sup_{{\kappa} > \kappa_{min}} |Y_{\kappa}(R)|$ and $\sup_{{\kappa} > \kappa_{min}} |Z_{\kappa}(R)|$ vanish for $R \rightarrow +\infty$ as well.

\section{Scattering wave functions in the vicinity of $k = 0^+$} \label{scat_wfs_k_zero}
We have demonstrated in Sec.(\ref{large_R_sec}) that scattering wave functions converge to their open radial interval limit
uniformly when $R \rightarrow +\infty$ for all linear momentum intervals $[k_d:+\infty[$, $k_d > 0$. 
However, uniform convergence is generally absent on $]0:+\infty[$, which is obvious when the set of positive energies wave functions
always contain one nodeless state for $R \rightarrow +\infty$, whose energy goes to zero.
It is thus necessary to devise the behavior of wave functions in the vicinity of $k = 0^+$. 
For this, considered linear momenta can always be chosen smaller than a positive $k_{\epsilon}$ momentum, which can be arbitrarily small
but has to be independent of $R$.
We will consider separately potentials with attractive and repulsive tails for $r \rightarrow +\infty$.

\subsection{Attractive tail}  \label{small_k_attractive}
Assuming $V_c < 0$, it is always possible to choose $R_1 > R_0$ large enough so that the remaining Coulomb + centrifugal
potential is negative $\forall r > R_1$. 
We can then define a fundamental ansatz solution of Eq.(\ref{Schrodinger_eq}) analogously to Sec.(\ref{large_k})
$\forall r > R_1$:
\begin{eqnarray}
&&\lambda_m(r) = \sqrt{k_m^2 + \frac{|V_c|}{r} - \frac{\ell(\ell+1)}{r^2}} \mbox{ , } 
\Lambda_m(r) = \int_{R_1}^{r} \lambda_m(r')~dr' \mbox{ , } \nonumber \\
&&u_{f_0}(r) = \lambda_m(r)^{-\frac{1}{2}} e^{\pm i \Lambda_m(r)} \mbox{ , } u_{f}(r) = u_{f_0}(r) [1 + \epsilon(r)], \label{uf_lambdas_small_k}
\end{eqnarray}
where $k_m$ is the linear momentum associated to the eigenstate of Eq.(\ref{Schrodinger_eq}), $u(k_m,r)$, $m \in \N^*$.

One derives the asymptotic relation yielded by $u(k_m,r)$ with a Neumann-Liouville power series in $r^{-\frac{1}{2}}$,
similarly as in Sec.(\ref{large_k}). Eq.(\ref{int_eq}) is used with $r_d = +\infty$ so that $\epsilon(r) \rightarrow 0$
for $r \rightarrow +\infty$. 
Using analytical expressions of $\lambda_m(r)$, $\Lambda_m(r)$ and $u_{f_0}(r)$ of Eq.(\ref{uf_lambdas_small_k}),
the equality $\epsilon(r) = O \left( r^{-\frac{1}{2}} \right)$ is procured for $r \rightarrow + \infty$
with partial integrations of Eq.(\ref{int_eq}). Hence :
\begin{eqnarray}
u(k_m,r) = \sqrt{\frac{2}{\pi}} \mathcal{N}_{k_m} \left(\frac{\lambda_m(r)}{k_m}\right)^{-\frac{1}{2}} \left[ \sin(\Lambda_m(r) + \delta_m)
 + O \left( r^{-\frac{1}{2}} \right) \right]
\mbox{ , } \forall r \geq R_1, \label{u_Ck_small_k}
\end{eqnarray}
where one has introduced the normalization $\mathcal{N}_{k_m}$, defined in Eq.(\ref{u_large_R_h_pm}), 
so that Eqs.(\ref{u_large_R_h_pm},\ref{u_Ck_small_k}) are consistent,
$\delta_m \in [0:2 \pi]$ is the phase shift associated to $u(k_m,r)$ and $O \left( r^{-\frac{1}{2}} \right)$ remains bounded for $k_m \rightarrow 0^+$.
One has $\delta_{m-1} - \delta_{m} \rightarrow 0$ when $R \rightarrow +\infty$ uniformly for $k_m \in ]0:k_{\epsilon}]$, $m \in \N^*$,
as phase shift is a uniformly continuous function of $k$ on $]0:k_{\epsilon}]$.

Normalization of $u(k_m,r)$ is defined in Eq.(\ref{u_n_km_norm}) and provides, along with Eq.(\ref{u_Ck_small_k}), the inequality:
\begin{eqnarray}
\frac{2}{\pi} \mathcal{N}_{k_m}^2 \int_{R_1}^{R} \lambda_m(r)^{-1} \left[ \sin^2(\Lambda_m(r) + \delta_m) + O \left( r^{-\frac{1}{2}} \right) \right]~dr 
\leq \frac{1}{k_m (k_{m} - k_{m-1})}. \label{ineq_norm_small_k}
\end{eqnarray}
App.(\ref{norm_equiv}) procures an inequality from Eq.(\ref{ineq_norm_small_k}) for $R$ sufficiently large:
\begin{eqnarray}
\frac{1}{2 \pi} \mathcal{N}_{k_m}^2 \int_{R_1}^{R} \lambda_m(r)^{-1}~dr \leq \frac{1}{k_m(k_{m} - k_{m-1})} \label{ineq_asymp_norm_small_k}.
\end{eqnarray}

The $k_{m} - k_{m-1}$ difference inverse is now majored.
Using $u(k_{m-1},R) = u(k_{m},R) = 0$ conditions, the fact that $k_{m-1}$ and $k_{m}$ are consecutive eigenmomenta
and Eq.(\ref{u_Ck_small_k}), one obtains:
\begin{eqnarray}
&&\sin(\Lambda_{m-1}(R) + \delta_{m-1}) + O(R^{-\frac{1}{2}}) = 0 \mbox{ , } \sin(\Lambda_{m}(R) + \delta_{m}) + O(R^{-\frac{1}{2}}) = 0 \nonumber \\
&\Rightarrow& \Lambda_{m-1}(R) = (n-1)\pi - \delta_{m-1} + O(R^{-\frac{1}{2}}) \mbox{ , } 
\Lambda_{m}(R) = n \pi - \delta_{m} + O(R^{-\frac{1}{2}}) \mbox{ , } n \in \N^* 
\nonumber \\
&\Rightarrow& \int_{R_1}^{R} (\lambda_{m}(r) - \lambda_{m-1}(r))~dr = \pi + \delta_{m-1} - \delta_{m} + O(R^{-\frac{1}{2}}).  \label{km_diff_small_k_1}
\end{eqnarray}
One derives from Eq.(\ref{km_diff_small_k_1}):
\begin{eqnarray}
(k_{m}^2 - k_{m-1}^2) \int_{R_1}^{R} \frac{dr}{\lambda_{m}(r) + \lambda_{m-1}(r)} = \pi + \delta_{m-1} - \delta_{m} 
+ O(R^{-\frac{1}{2}}) \label{km_diff_small_k_2},
\end{eqnarray} 
where one has applied the conjugate $\lambda_{m}(r) + \lambda_{m-1}(r)$ of $\lambda_{m}(r) - \lambda_{m-1}(r)$ in Eq.(\ref{km_diff_small_k_1}) 
in order to have $k_{m}^2 - k_{m-1}^2$ appeared in Eq.(\ref{km_diff_small_k_2}).
One has $\lambda_{m-1}(r) \geq 0$ and $\pi + \delta_{m-1} - \delta_{m} + O(R^{-\frac{1}{2}}) \geq \frac{\pi}{2}$ for $R$ large enough,
so that Eq.(\ref{km_diff_small_k_2}) provides:
\begin{eqnarray}
\frac{1}{k_{m} - k_{m-1}} \leq \frac{4 k_m}{\pi} \int_{R_1}^{R} \lambda_m(r)^{-1}~dr \label{km_diff_small_k_3}
\end{eqnarray} 
Combining Eqs.(\ref{ineq_asymp_norm_small_k},\ref{km_diff_small_k_3}),
the normalization factor $\mathcal{N}_{k_m}^2$ can be seen to be uniformly majored for $k_m \in ]0:k_{\epsilon}[$ when $R \rightarrow +\infty$:
\begin{eqnarray}
\mathcal{N}_{k_m}^2 \leq M \label{norm_ineq_small_k_attractive},
\end{eqnarray} 
with $M > 0$ independent of $k_m$, $k_{\epsilon}$ and $R$.

If $V_c = \ell = 0$, one has seen in Sec.(\ref{large_R_sec}) that Eq.(\ref{norm_large_R}) is available uniformly for $k \in ]0:+\infty]$,
so that Eq.(\ref{norm_ineq_small_k_attractive}) can also be utilized in this case.

\subsection{Repulsive tails} \label{small_k_repulsive}
Low-energy scattering states of a potential with a repulsive tail bear a turning point $r_t(k)$ at large distance:
\begin{eqnarray}
r_t(k) = \frac{V_c + \sqrt{V_c^2 + 4 \ell(\ell+1) k^2}}{2 k^2}. \label{rt_def_repulsive}
\end{eqnarray}
It prevents a WKB approximation similar to Eq.(\ref{uf_lambdas_small_k}) from being devised. 
However, low-energy box scattering states bear the property to become very close to regular Coulomb wave functions
$F_{\ell \eta}(kr)$ up to one constant when $k \rightarrow 0^+$, which will be demonstrated in this section.
As the set of regular Coulomb wave functions can be shown to be complete, using complex integration techniques 
and analytic properties of confluent hypergeometric functions 
(see Ref.\cite{Mukunda} for the Coulomb attractive case and App.(\ref{cwf_completeness}) for its extension to repulsive case),
it will be possible to handle the contribution of low-energy box scattering states accumulating at $k = 0^+$ properly. 
Non-Coulombic, but repulsive tails, for which $\ell(\ell+1) > 0$, are considered in the same manner, $F_{\ell \eta}(kr)$ 
reducing to Riccati-Bessel functions $\hat{j}_{\ell}(kr)$ with all stated results remaining available.

One considers $0 < r < r_t(k)$ fixed, $r_t(k)$ defined in Eq.(\ref{rt_def_repulsive}) and $k < k_{\epsilon}$ an eigenmomentum of Eq.(\ref{Schrodinger_eq}).
There is no loss of generality therein as it is always possible to take $k_{\epsilon}$ small enough so that it is the case $\forall k < k_{\epsilon}$. 
Let us write the equivalents of $F_{\ell \eta} (kr)$ and $G_{\ell \eta} (kr)$ for $k \rightarrow 0^+$ and $r$ fixed,
omitting factors independent of $r$ and $k$ \cite{Abramowitz_Stegun}:
\begin{eqnarray}
&&F_{\ell \eta} (kr) \sim {\eta}^{-\frac{1}{2}} r^{\frac{1}{4}} e^{-\pi \eta} i_{2 \ell + \frac{1}{2}} (2 \sqrt{V_c r}) \mbox{ , }
G_{\ell \eta} (kr) \sim {\eta}^{-\frac{1}{2}} r^{\frac{1}{4}} e^{\pi \eta} k_{2 \ell + \frac{1}{2}} (2 \sqrt{V_c r}) \mbox{ , } (V_c > 0) \mbox{ , } \nonumber \\
&&F_{\ell \eta} (kr) \sim (kr)^{\ell+1} \mbox{ , } G_{\ell \eta} (kr) \sim (kr)^{-\ell} \mbox{ , } (V_c = 0)
\label{FG_equivalents},
\end{eqnarray}
where $i_{2 \ell + \frac{1}{2}}(x)$ and $k_{2 \ell + \frac{1}{2}}(x)$ are the modified Riccati-Bessel functions
respectively regular at $x = 0$ and $x \rightarrow +\infty$.

If $r \geq R_0$, $u(r)$ is a linear combination of regular and irregular Coulomb wave functions:
\begin{eqnarray}
u(k,r) = \mathcal{C}(k) \left[ F_{\ell \eta} (kr) + A(k) G_{\ell \eta} (kr) \right] \label{u_after_R0},
\end{eqnarray}
where $\mathcal{C}(k) \in \R$ is a normalization constant and $A(k) \in \R$ a matching constant.
One will show that \\ $A(k) G_{\ell \eta} (k R_0) = O(F_{\ell \eta} (k R_0))$.
From Eqs.(\ref{FG_equivalents},\ref{u_after_R0}), 
supposing the opposite would imply the $G_{\ell \eta} (kr)$ component of $u(k,r)$ to be dominant at $r=R_0$ for $k \rightarrow 0^+$.
Then, $u(k,r) \propto G_{\ell \eta} (kr)$ asymptotically for $k \rightarrow 0^+$ and $R_0 \leq r \leq R_1$, with $R_1 \rightarrow + \infty$.
This is impossible because the considered Hamiltonian in open radial interval bears no bound state at $k = 0$,  
so that $|u_0(r)| \rightarrow +\infty$ for $r \rightarrow +\infty$, with $u_0(r)$
the regular solution of Eq.(\ref{Schrodinger_eq}) calculated at $k = 0$ and arbitrarily normalized.

This and the fact that $F_{\ell \eta} (kr)$ and $G_{\ell \eta} (kr)$ have comparable amplitudes for $r > r_t(k)$ \cite{Abramowitz_Stegun} 
imply that norms and energies of box scattering states defined with Eq.(\ref{Schrodinger_eq}) and those defined from pure Coulomb + centrifugal potentials
will be the same up to a value which can be made arbitrarily small $\forall k \in [0:k_{\epsilon}]$ and $k_{\epsilon}$ small enough. 
Thus, $\forall k_m \in ]0:k_{\epsilon}]$, $m \in N^*$, for $R$ large enough, $k_{\epsilon}$ small enough and $R_0 \leq r < r_t(k_{\epsilon})$:
\begin{eqnarray}
|u(k_m,r)| < A~u_c({k'}_m,r) \mbox{ , } |k_{m} - k_{m-1}| < 2 (k'_{m} - k'_{m-1})  \label{ineq_small_k_repulsive},
\end{eqnarray}
where $u(k_m,r)$, $m \in \N^*$ is a box discretized scattering state of Eq.(\ref{Schrodinger_eq}),
$u_c({k'}_m,r)$ is a box discretized scattering state but whose Hamiltonian is one of pure Coulomb + centrifugal parts,
proportional to $F_{\ell \eta_m} ({k'}_m r)$, whose energy differ by that of $u(k_m,r)$ by an arbitrarily small amount,
and $A > 0$ is independent of $k_m$, $k'_m$ and $R$ but can be dependent of $r$. 

If $r \leq R_0$, $u(k,r)$ verifies for $k \rightarrow 0^+$:
\begin{eqnarray}
u(k,r) = \mathcal{D}(k) (u_0(r) + o(1)) \label{u_bef_R0_1},
\end{eqnarray}
where $\mathcal{D}(k)$ is a normalization factor. 
As $u(k,R_0)$ is also provided by Eq.(\ref{u_after_R0}), one can match Eqs.(\ref{u_after_R0},\ref{u_bef_R0_1}) at $r=R_0$,
so that Eq.(\ref{FG_equivalents}) and the fact that $A(k) G_{\ell \eta} (k R_0) = O(F_{\ell \eta} (k R_0))$
provides for $k \rightarrow 0^+$ the asymptotic value of $\mathcal{D}(k)$:
\begin{eqnarray}
\mathcal{D}(k) &\sim& \mathcal{C}(k)~D~{\eta}^{-\frac{1}{2}} e^{-\pi \eta} \mbox{ , } (V_c > 0) \mbox{ , } \nonumber \\ 
               &\sim& \mathcal{C}(k)~D~k^{\ell+1} \mbox{ , } (V_c = 0), \label{Dk_def}
\end{eqnarray}
with $D \in \R$ bounded for $k \rightarrow 0^+$. Thus, for $k \rightarrow 0^+$:
\begin{eqnarray}
u(k,r) &=& \mathcal{C}(k)~D~{\eta}^{-\frac{1}{2}} e^{-\pi \eta} (u_0(r) + o(1)) \mbox{ , } (V_c > 0) \mbox{ , } \nonumber \\
       &=& \mathcal{C}(k)~D~k^{\ell+1}  (u_0(r) + o(1)) \mbox{ , } (V_c = 0) \mbox{ , } \nonumber \\
       &=& f(r)~\mathcal{C}(k) F_{\ell \eta} (kr) \label{u_bef_R0_2},
\end{eqnarray}
where $f(r)$ is bounded for $k \rightarrow 0^+$ from Eq.(\ref{FG_equivalents}).
Hence, the left-hand side of Eq.(\ref{ineq_small_k_repulsive}) also applies for $r \leq R_0$.

\section{Asymptotic treatment of bound states of potentials with attractive Coulomb tail} \label{large_n}
Bound states in open radial interval in our class of Hamiltonians are in finite number except if $v(r)$
behaves as a Coulomb attractive potential for $r \rightarrow +\infty$, hence with $V_c < 0$ in Eq.(\ref{v_asymp_inf}), 
which is now assumed for this section.
As the sum of bound states in Eq.(\ref{formal_conv_series}) becomes an infinite series for $R \rightarrow +\infty$,
with all energies accumulating at zero energy,
it is necessary to devise the asymptotic behavior of bound states for large principal quantum number to study its convergence properties.
It is convenient to use $\kappa_n = -i k_n$, with $n \in \N$ the number of nodes and $k_n$ the complex linear momentum of the bound eigenstate $u_n(r)$,
so that $\kappa_n \rightarrow 0^+$ for $n \rightarrow +\infty$. $R$ is naturally demanded to be larger than $R_0$ 
to have the non-local part of Eq.(\ref{Schrodinger_eq}) disappeared. 
Owing to Eq.(\ref{Schrodinger_eq}), one defines the Coulomb turning point $r_t^{(n)}$ of the bound eigenstate $u_n(r)$ for $\kappa_n \rightarrow 0$:
\begin{eqnarray}
r_t^{(n)} = \frac{|V_c| + \sqrt{V_c^2 - 4 \ell(\ell+1) \kappa_n^2}}{2 \kappa_n^2} = \frac{|V_c|}{\kappa_n^2} + O(1). \label{rt_def}
\end{eqnarray}
To differentiate wave functions defined in open radial interval and with box boundary conditions, one will write $\widetilde{u}_n(r)$
in open radial interval and $u_n(r)$ for box wave functions.
One then defines the fundamental solution ansatz of Eq.(\ref{Schrodinger_eq}) suitable for bound states of large principal quantum number $n$:
\begin{eqnarray}
&&\lambda_n(r) = \sqrt{\frac{|V_c|}{r} - \frac{\ell(\ell+1)}{r^2} - \kappa_n^2} \mbox{ , } 
\Lambda_n(r) = \int_{r_d}^{r} \lambda_n(r')~dr' \mbox{ , } u_{f}(r) = \lambda_n(r)^{-\frac{1}{2}} e^{\pm i \Lambda_n(r)} [1 + \epsilon_n(r)], 
\label{lambdas_uf_large_n}
\end{eqnarray}
with $r_d > R_0$, chosen large enough but independently of $n$, so that the square root argument in $\lambda_n(r)$ is always strictly positive
$\forall r \in [r_d:r_t^{(n)}[$ $\forall n > N$, $N \in \N$.
Similarly to Sec.(\ref{small_k_attractive}), $\widetilde{u}_n(r)$ is shown to verify the following asymptotic relation
by way of Eq.(\ref{lambdas_uf_large_n}) for $n \rightarrow +\infty$:
\begin{eqnarray}
\widetilde{u}_n(r) = C_n~\lambda_n(r)^{-\frac{1}{2}} \left[ \sin(\Lambda_n(r) + \delta_n) + O \left( r^{-\frac{1}{2}} \right) \right]
\mbox{ , } r \leq r_e^{(n)} \label{u_Cn_large_n},
\end{eqnarray}
where $r_e^{(n)} = \frac{r_t^{(n)}}{2}$, value introduced to avoid the divergences arising close to the turning point $r_t^{(n)}$,
$O \left( r^{-\frac{1}{2}} \right)$ remains bounded for $n \rightarrow +\infty$,
$C_n \in \R$ is a normalization constant and $\delta_n \in [0:2 \pi]$ plays the role of a phase shift for $\widetilde{u}_n(r)$.
As $C_n^{-1} \widetilde{u}_n(r)$ has a finite limit for $n \rightarrow +\infty$ for $r \geq r_d$ fixed from Eq.(\ref{u_Cn_large_n})
and as $u_n(0) = 0$ $\forall n \in \N$, $C_n^{-1} \widetilde{u}_n(r) = O(1)$ $\forall r \geq 0$ fixed and $n \rightarrow +\infty$.
As $\widetilde{u}_n(r)$ is normalized, one can write:
\begin{eqnarray}
\int_{0}^{r_d} \widetilde{u}_n^2(r')~dr' + \int_{r_d}^{r_e^{(n)}} \!\!\!\! \widetilde{u}_n^2(r')~dr' \leq 1. \label{un_truncated_norm}
\end{eqnarray}
The first integral of Eq.(\ref{un_truncated_norm}) divided by $C_n^2$ has a finite limit for $n \rightarrow +\infty$ as $r_d$ is independent of $n$.
Hence, the first integral of Eq.(\ref{un_truncated_norm}) is $O(C_n^2)$ for $n \rightarrow +\infty$.
Using Eqs.(\ref{rt_def},\ref{lambdas_uf_large_n},\ref{u_Cn_large_n}) and App.(\ref{norm_equiv}), 
one obtains for the second integral of Eq.(\ref{un_truncated_norm}) with $n \rightarrow +\infty$:
\begin{eqnarray}
\int_{r_d}^{r_e^{(n)}} \!\!\!\! \widetilde{u}_n^2(r')~dr' 
&=& C_n^2~\int_{r_d}^{r_e^{(n)}} \!\!\!\! \lambda_n(r)^{-1} \left[ \sin^2 \left( \Lambda_n(r) + \delta_n \right) + O \left( r^{-\frac{1}{2}} \right) \right]~dr 
\nonumber \\
&\sim& \frac{C_n^2}{2} \int_{r_d}^{r_e^{(n)}} \!\!\!\! \lambda_n(r)^{-1}~dr
\sim \frac{C_n^2~\kappa_n^{-3}}{2} 
       \int_{\kappa_n^2 r_d}^{\kappa_n^2 r_e^{(n)}}~\left( \frac{|V_c|}{t} - 1 - \frac{\ell(\ell+1)}{t^2} \kappa_n^2 \right)^{-\frac{1}{2}}~dt \nonumber \\
&=& O(C_n^2~\kappa_n^{-3}).
\label{second_integral_norm_large_n}
\end{eqnarray}
Owing to boundedness of $\widetilde{u}_n(r)$ divided by $C_n$ for $n \rightarrow +\infty$ and $r > 0$ fixed,
Eqs.(\ref{un_truncated_norm},\ref{second_integral_norm_large_n}) provide: 
\begin{eqnarray}
C_n^2 = O(\kappa_n^{3}) \mbox{ , } \widetilde{u}_n(r) = O \left( {\kappa_n}^{\frac{3}{2}} \right) \mbox{ , } r \geq 0 \mbox{ fixed , } n \rightarrow +\infty. 
\label{un_kappa_n}
\end{eqnarray}
The following expansion is standard from quantum defect theory \cite{Seaton}:
\begin{eqnarray}
\kappa_n = O(n^{-1}) \mbox{ , } n \rightarrow +\infty. \label{kappa_QDT}
\end{eqnarray}
Eqs.(\ref{un_kappa_n},\ref{kappa_QDT}) then provide:
\begin{eqnarray}
\widetilde{u}_n(r) = O(n^{-\frac{3}{2}}) \mbox{ , } r \geq 0 \mbox{ fixed , } n \rightarrow +\infty. \label{u_large_n}
\end{eqnarray}
The case of Hamiltonian defined with box boundary conditions will be shown to be very similar for $R \rightarrow +\infty$, 
as long as $n \leq N_b(R)$, with $N_b(R)$ the number of bound states with box boundary conditions, diverging for $R \rightarrow +\infty$. 
For $n \leq N_b(R)$, $u_n(R) = 0$,
so that consideration of the number of nodes of wave functions implies that $\kappa_{n+1} < \kappa_n(R) < \kappa_n$, with $\kappa_n$ and $\kappa_{n+1}$ 
defined in open radial interval, and $\kappa_n(R)$ defined with box boundary conditions. Eq.(\ref{u_large_n}) thus remains
valid for box bound states, as long as $n \rightarrow +\infty$ along with $R$ so that $n \leq N_b(R)$ and $r_e^{(n)} \leq R$.
In case $r_e^{(n)} > R$, Eq.(\ref{rt_def}) and the fact that $\kappa_n(R) \sim \kappa_n$  for $n \rightarrow +\infty$ and $n \leq N_b(R)$
imply that $\kappa_n^2 R = O(1)$ therein.
$r_e^{(n)}$ has to be replaced by $R$ in Eq.(\ref{un_truncated_norm}) so that using the same method as in open radial interval, one obtains
$u_n(r) = O \left( {\kappa_n}^{\frac{3}{2}} \right)$ for $n \rightarrow +\infty$ and $n \leq N_b(R)$. 
Eq.(\ref{u_large_n}) is thus still valid in this case.
Note that $n$, and hence $R$, have to be taken large enough so that wave functions $u_{n'}(r)$ are very close for $r \in [0:R_0]$ $\forall n' > n$,
in order for the influence of the non-local potential of Eq.(\ref{Schrodinger_eq}) to be negligible.
This condition is necessary, as the number of nodes of wave functions is uncorrelated with energy for the general non-local potential.

\section{Completeness relation of bound and scattering wave functions in the open radial interval} \label{compl_rel_open_space}
The limit of Eq.(\ref{formal_conv_series}) for $R \rightarrow +\infty$ will be shown to be, as can be expected:
\begin{eqnarray}
S(r,r') &=& \sum_{n=0}^{+\infty} \widetilde{u}_n(r) \widetilde{u}_n(r') 
         + \int_{0}^{+\infty} \left[ \widetilde{u}(k,r) \widetilde{u}(k,r')
         - \frac{2}{\pi} \hat{j}_{\ell}(k r) \hat{j}_{\ell}(k r') \right]~dk, \label{formal_conv_int}
\end{eqnarray}
where $\{\widetilde{u}_n(r)\}_{n \in \N}$ is the set of bound states of the Hamiltonian of Eq.(\ref{Schrodinger_eq}) defined in open radial interval
and $\{\widetilde{u}(k,r)\}_{k \in ]0:+\infty[}$ its set of scattering states, reserving notation without tilde symbol for states defined in a finite interval
of radius $R$, as in Sec.(\ref{large_n}). Scattering states are normalized with Dirac delta normalization.
The first step is to show that series and integral in Eq.(\ref{formal_conv_int})
converge $\forall (r,r') \in [0:+\infty[^2$. Convergence of integral in $k \rightarrow 0^+$ is guaranteed by boundedness of scattering wave functions
therein for fixed $(r,r')$. If $V_c < 0$ or $\ell = V_c = 0$, this property is a direct consequence of Eq.(\ref{u_Ck_small_k}),
applied with $k_m = k$ and $\mathcal{N}_{k_m} = 1$, equivalent to Dirac delta normalization. $\widetilde{u}(k,r)$ can be seen to have therein a finite limit 
for $k \rightarrow 0^+$ $\forall r \geq R_1$ (see Sec.(\ref{small_k_attractive}) for definitions and notations). 
Boundedness $\forall r \geq 0$ comes forward due to $\widetilde{u}(k,0) = 0$ equality $\forall k > 0$.
If $V_c > 0$ or $V_c = 0, \ell > 0$, Dirac delta normalization is equivalent to have $\mathcal{C}(k)^2 (1 + A(k)^2) = \frac{2}{\pi}$
in Eq.(\ref{u_after_R0}). For $k \rightarrow 0^+$, one obtains $\mathcal{C}(k) \sim \sqrt{\frac{2}{\pi}}$ (see Sec.(\ref{small_k_repulsive})).
Boundedness of $\widetilde{u}(k,r)$ $\forall r \geq 0$ arises similarly as in the previous case.
Problems of convergence, however, would occur in $k = 0$ in the complex $k$-plane,
as the Green's function of Hamiltonians with Coulomb asymptotic possesses an essential singularity therein \cite{Newton_book}.
Behavior of Eq.(\ref{formal_conv_int}) at $k \rightarrow +\infty$ is dealt with
as in Sec.(\ref{compl_rel_diff_box}). Eq.(\ref{u_large_k}) and Dirac delta normalization of $\widetilde{u}(k,r)$ imply that, for $k \rightarrow +\infty$:
\begin{eqnarray}
\widetilde{u}(k,r) \widetilde{u}(k,r') - \frac{2}{\pi} \hat{j}_{\ell}(k r) \hat{j}_{\ell}(k r')
= \! \frac{\mathcal{V}(r) - \mathcal{V}(r')}{2 \pi k}  \sin [k(r-r')]
- \frac{\mathcal{V}(r) + \mathcal{V}(r')}{2 \pi k}  \sin [k(r+r') - \ell \pi]
+ O \left( \frac{1}{k^2} \right). \label{int_term_asymp} 
\end{eqnarray}
As Eq.(\ref{int_term_asymp}) possesses the same characteristics as Eq.(\ref{series_term_asymp}),
normal convergence of the integral of Eq.(\ref{formal_conv_int}) at $k \rightarrow +\infty$ with respect to $(r,r') \in [0:R_t]^2$, $R_t > 0$,
is achieved by way of a partial integration playing the role of Abel transformation in Sec.(\ref{compl_rel_diff_box}).
Normal convergence of the series in Eq.(\ref{formal_conv_int}) is provided by Eq.(\ref{u_large_n}), so that $S(r,r')$ is continuous 
$\forall (r,r') \in [0:+\infty[^2$.

The difference $S(r,r') - S_R(r,r')$ issued from Eqs.(\ref{formal_conv_series},\ref{formal_conv_int}) has to be proved to vanish for $R \rightarrow +\infty$.
It is rewritten as:
\begin{eqnarray}
&&S(r,r') - S_R(r,r') \nonumber \\
&=& -\sum_{m=1}^{+\infty} \left[ \left( \mathcal{N}_{k_m}^2 - 1 \right) \widetilde{u}(k_m,r) \widetilde{u}(k_m,r') (k_{m} - k_{m-1})
- \left( B_{\kappa_m}^2 - \frac{2}{\pi} \right) \hat{j}_{\ell}(\kappa_m r) \hat{j}_{\ell}(\kappa_m r') (\kappa_{m} - \kappa_{m-1}) \right]  \nonumber \\
&+& \sum_{n=0}^{+\infty} [\widetilde{u}_n(r) \widetilde{u}_n(r') - u_n(r) u_n(r')] 
+ \int_{0}^{+\infty} \left[ \widetilde{u}(k,r) \widetilde{u}(k,r') - \frac{2}{\pi} \hat{j}_{\ell}(k r) \hat{j}_{\ell}(k r')\right]~dk \nonumber \\
&-& \sum_{m=1}^{+\infty} \left[ \widetilde{u}(k_m,r) \widetilde{u}(k_m,r') (k_{m} - k_{m-1})
- \frac{2}{\pi} \hat{j}_{\ell}(\kappa_m r) \hat{j}_{\ell}(\kappa_m r') (\kappa_{m} - \kappa_{m-1}) \right]  \label{S_diff}.
\end{eqnarray}
The series of Eq.(\ref{formal_conv_series}) has been separated in two converging parts in Eq.(\ref{S_diff})
(see Eqs.(\ref{Cm_Bm_two_pi},\ref{norm_large_R}) and Sec.(\ref{compl_rel_diff_box})). 
The bound states $u_n(r)$ defined with box boundary conditions are finite in number, so that $u_n(r) = 0$ by definition
for $n > N_b(R)$, $N_b(R)$ being the number of box bound states. In case $v(r)$ has an attractive Coulomb tail, 
using Eq.(\ref{u_large_n}) and boundedness of bound states,
all functions $u_n(r)$ and $\widetilde{u}_n(r)$, $n \in \N$, can be majored in absolute value by a constant independent of $R$ and $n$ 
divided by $1+n^{\frac{3}{2}}$ for $r \geq 0$ fixed. Normal convergence of the bound states series with respect to $R$ comes forward.
The studied series reduces to a finite sum in other cases. The series thus goes to zero with $R \rightarrow +\infty$ as each of its terms obviously does.

The two last terms of Eq.(\ref{S_diff}), a series and an integral, resemble a Riemann sum and its integral limit for infinite number of points.
The fact that the sum is already a series for finite $R$ and that the integral is improper, however, renders impossible the direct application of Riemann integral
fundamental theorem. Demonstrating that the difference of the considered series and integral vanishes when $R \rightarrow +\infty$ is performed 
in App.(\ref{Riemann_improper}).

The first infinite series of Eq.(\ref{S_diff}), denoted as $S_l(r,r')$, is equal to:
\begin{eqnarray}
S_l(r,r') =  \sum_{m=1}^{+\infty} \left[ \left( \mathcal{N}_{k_m}^2 - 1 \right) \widetilde{u}(k_m,r) \widetilde{u}(k_m,r') (k_{m} - k_{m-1})
-  \left( B_{\kappa_m}^2 - \frac{2}{\pi} \right) \hat{j}_{\ell}(\kappa_m r) \hat{j}_{\ell}(\kappa_m r') (\kappa_{m} - \kappa_{m-1}) \right]. \label{Sl_value}
\end{eqnarray}
Let $k_{\epsilon} > 0$ and $m_{\epsilon} \in \N^*$ be an integer verifying $k_{m_{\epsilon}} \rightarrow k_{\epsilon}$ for $R \rightarrow +\infty$.
As $u(k_{m_{\epsilon}},r)$ and $\hat{j}_{\ell}(\kappa_{m_{\epsilon}} r)$ have the same number of nodes $\forall R > 0$ (see Sec.(\ref{compl_rel_diff_box})),
$\kappa_{m_{\epsilon}}$ can never be arbitrarily small for $R \rightarrow +\infty$. 
Eq.(\ref{kn_eq_large_R}) can then be applied to $k_{m_{\epsilon}}$ and $\kappa_{m_{\epsilon}}$,
so that $\kappa_{m_{\epsilon}} \rightarrow k_{\epsilon}$ for $R \rightarrow +\infty$. 
We will now always implicitly suppose that the limit $R \rightarrow +\infty$ has been effected before proceeding with $k_{\epsilon} \rightarrow 0$.

One has $S_l(r,r') = S^{\epsilon -}_{l}(r,r') + S^{\epsilon +}_{l}(r,r')$, whose respective general term is the same as $S_l(r,r')$, with $m < m_{\epsilon}$
for $S^{\epsilon -}_{l}(r,r')$ and $m \geq m_{\epsilon}$ for $S^{\epsilon +}_{l}(r,r')$. $S^{\epsilon +}_{l}(r,r')$
is demonstrated to vanish for $R \rightarrow +\infty$ in App.(\ref{norm_diff_series}). It is thus trivially true for $k_{\epsilon} \rightarrow 0^+$. 
We now study $S^{\epsilon -}_{l}(r,r')$
by separating its Bessel and non-Bessel parts, which is permitted as $m_{\epsilon}$ is finite $\forall R > 0$:
\begin{eqnarray}
|S^{\epsilon -}_{l}(r,r')| &\leq& 
\left| \sum_{m=1}^{m_{\epsilon-1}} \left( \mathcal{N}_{k_m}^2 - 1 \right) \widetilde{u}(k_m,r) \widetilde{u}(k_m,r') (k_{m} - k_{m-1}) \right|
\nonumber \\
&+& \left|
\sum_{m=1}^{m_{\epsilon-1}} \left( B_{\kappa_m}^2 - \frac{2}{\pi} \right) \hat{j}_{\ell}(\kappa_m r) \hat{j}_{\ell}(\kappa_m r') (\kappa_{m} - \kappa_{m-1})
\right|
\label{Sl_minus_ineq}.
\end{eqnarray}
If $V_c < 0$ or $\ell = V_c = 0$, one can use results of Sec.(\ref{small_k_attractive}), 
so that Eq.(\ref{norm_ineq_small_k_attractive}) imply for the first sum of Eq.(\ref{Sl_minus_ineq}), $R \rightarrow +\infty$:
\begin{eqnarray}
\left| \sum_{m=1}^{m_{\epsilon-1}} \left( \mathcal{N}_{k_m}^2 - 1 \right) \widetilde{u}(k_m,r) \widetilde{u}(k_m,r') (k_{m} - k_{m-1}) \right|
&\leq& \sum_{m=1}^{m_{\epsilon-1}} \left( \mathcal{N}_{k_m}^2 + 1 \right) |\widetilde{u}(k_m,r) \widetilde{u}(k_m,r')| (k_{m} - k_{m-1}) \nonumber \\
&\leq& M \sum_{m=1}^{m_{\epsilon-1}} (k_{m} - k_{m-1}) \sim M k_{\epsilon} , \label{Sl_minus_ineq_attractive}
\end{eqnarray}
where $M > 0$ is independent of $R$ and $k_{\epsilon}$. 
Eq.(\ref{Sl_minus_ineq_attractive}) clearly vanishes for $k_{\epsilon} \rightarrow 0$, 
result also available for the second sum of Eq.(\ref{Sl_minus_ineq}) if $\ell = 0$.
If the potential is repulsive in its asymptotic region, one is in the conditions stated in Sec.(\ref{small_k_repulsive}), 
so that Eq.(\ref{ineq_small_k_repulsive}) accounts for the following inequality for $k_{\epsilon}$ small enough and $R$ large enough:
\begin{eqnarray}
\left| \sum_{m=1}^{m_{\epsilon-1}} \left( \mathcal{N}_{k_m}^2 - 1 \right) \widetilde{u}(k_m,r) \widetilde{u}(k_m,r') (k_{m} - k_{m-1}) \right|
\leq \sum_{m=1}^{m_{\epsilon-1}} (A u_c({k'}_m,r) u_c({k'}_m,r') + B) ({k'}_{m} - {k'}_{m-1}), \nonumber \\
\leq A \sum_{m=1}^{m_{\epsilon-1}} u_c({k'}_m,r) u_c({k'}_m,r') ({k'}_{m} - {k'}_{m-1}) + 2 B k_{\epsilon} \label{Sl_minus_ineq_repulsive}
\end{eqnarray}
where $A > 0$ and $B > 0$ are independent of $R$ and $k_{\epsilon}$, and notations are the same as in Sec.(\ref{small_k_repulsive}).
Eq.(\ref{Sl_minus_ineq_repulsive}) is available as well for the second sum of Eq.(\ref{Sl_minus_ineq}) if $\ell(\ell+1) > 0$. 
Let us suppose that the left-hand side of Eq.(\ref{Sl_minus_ineq_repulsive}) does not vanish for $k_{\epsilon} \rightarrow 0$,
assuming also for the moment $\ell = 0$.
In Eq.(\ref{Sl_minus_ineq_repulsive}), the sum involving $u_c(k'_m,r)$ functions, eigenstates of pure Coulomb + centrifugal potential, does not vanish as well. 
Thus, using $u_c(k'_m,r)$ functions instead of $u_n(r)$ and $u(k,r)$ states in Eq.(\ref{S_diff}), and denoting as $S_c(r,r')$
the $S(r,r')$ function of Eq.(\ref{S_diff}) which they generate, one obtains $S_c(r,r') \neq 0$ 
as all terms in Eq.(\ref{S_diff}) vanish for $R \rightarrow +\infty$ except the right-hand side of Eq.(\ref{Sl_minus_ineq_repulsive}).
This result is contradictory with completeness of $\widetilde{u}_c(k,r)$ and $\hat{j}_{\ell}(k r)$ 
functions, demonstrated independently in App.(\ref{cwf_completeness}). 
Indeed, with $R_t > 0$ fixed so that $S_c(r,r')$  is not identically equal to zero in $[0:R_t]^2$, one obtains $\forall r < R_t$:
\begin{eqnarray}
\int_{0}^{R_t} S_c(r,r')^2~dr' &=& \int_{0}^{+\infty} \widetilde{u}_c(k,r) \int_{0}^{R_t} S_c(r,r')~\widetilde{u}_c(k,r')~dr'~dk \nonumber \\
         &-& \frac{2}{\pi} \int_{0}^{+\infty} \hat{j}_{\ell}(k r) \int_{0}^{R_t} S_c(r,r')~\hat{j}_{\ell}(k r')~dr'~dk
= S_c(r,r) - S_c(r,r) = 0,  \label{Sl_contradiction}
\end{eqnarray}
where all integrals inversions are allowed due to normal convergence of integral in Eq.(\ref{formal_conv_int}) 
at $k \rightarrow +\infty$ with respect to $(r,r') \in [0:R_t]^2$.
If $\ell > 0$, assuming $V_c < 0$ implies that the Bessel sum of $S^{\epsilon -}_{l}(r,r')$ in Eq.(\ref{Sl_minus_ineq}) vanishes for $k_{\epsilon} \rightarrow 0$.
The remaining cases, with $\ell > 0$ and $V_c > 0$, can then be treated similarly as for $\ell = 0$.

Thus, Eqs.(\ref{Sl_minus_ineq_repulsive},\ref{Sl_contradiction}) imply that $S^{\epsilon -}_{l}(r,r')$ goes to zero for $k_{\epsilon} \rightarrow 0$, 
leading to $S_l(r,r') \rightarrow 0$ for $R \rightarrow +\infty$.
The equality $S(r,r') = 0$, $\forall (r,r') \in [0:+\infty[^2$ comes forward.
Completeness of $\{\widetilde{u}_n(r)\}_{n \in \N}$ and $\{\widetilde{u}(k,r)\}_{k \in ]0:+\infty[}$ is achieved similarly as in Eq.(\ref{Sl_contradiction}), 
with an arbitrary function $f(r)$ bearing a Fourier-Bessel transform:
\begin{eqnarray}
&&\int_{0}^{+\infty} f(r')~S(r,r')~dr' = 0
= \sum_{n=0}^{+\infty} \widetilde{u}_n(r) \int_{0}^{+\infty} f(r')~\widetilde{u}_n(r')~dr' \nonumber \\
         &+& \int_{0}^{+\infty} \left[ \widetilde{u}(k,r) \int_{0}^{+\infty} f(r')~\widetilde{u}(k,r')~dr'
         - \frac{2}{\pi} \hat{j}_{\ell}(k r) \int_{0}^{+\infty} f(r')~\hat{j}_{\ell}(k r')~dr'~dk \right]  \nonumber \\
&\Rightarrow& f_{j}(r) = \sum_{n=0}^{+\infty} \widetilde{u}_n(r) \int_{0}^{+\infty} f(r')~\widetilde{u}_n(r')~dr'
         + \int_{0}^{+\infty} \widetilde{u}(k,r) \int_{0}^{+\infty} f(r')~\widetilde{u}(k,r')~dr' \label{f_expansion},
\end{eqnarray}
where $f_{j}(r)$ is the value of the Fourier-Bessel expansion of $f$ in $r$.
Convergence and inversions of series and integral in Eq.(\ref{f_expansion}) are possible 
with $f(r)$ decreasing sufficiently fast for $r \rightarrow + \infty$,
as it implies normal convergence of series and integrals with respect to all parameters.
As in the discrete case, $f_{j}(r) = \lim_{\delta \rightarrow 0} \frac{f(r+\delta) + f(r-\delta)}{2}$.

\section{Generalizations}

\subsection{Coulomb divergence in $r \rightarrow 0$} \label{Coulomb_div_pot_zero}
In Sec.(\ref{large_k}), the Bessel-WKB approximation could be applied for $v(r)$ finite in $r=0$
and then seen available afterward for $v(r)$ therein integrable. The important case of pure Coulomb potential,
for which $v(r) \propto r^{-1}$ for $r \rightarrow 0$, however, has not been covered.
An obvious remedy is to use Coulomb wave functions instead of Bessel functions for the WKB approximation in Eq.(\ref{exact_function}), 
with $v(r)$ is replaced by $v_0(r) = v(r) - \frac{V_c}{r}$ in Sec.(\ref{large_k}).
One can check that this generates an additional term in Eq.(\ref{F_def}), 
equal to $\left( \frac{1}{r} - \frac{k^2 - v_0(r)}{k \Lambda(r)} \right) V_c$. It poses no problem as it is $O(k^{-2})$
for $k \rightarrow +\infty$, $O(r^{-1})$ for $r \rightarrow 0$, $k$ fixed and $O(r^{-2})$ for $r \rightarrow +\infty$, $k$ fixed as well. 
The Coulomb-WKB approximation is obtained following the same method
as in Sec.(\ref{large_k}):
\begin{eqnarray}
&&u(r) = C_k F_{\ell \eta} (kr) - C_k \frac{\mathcal{V}_0(r)}{2k} F'_{\ell \eta} (kr) + O \left( \frac{C_k}{k^2} \right) 
       \mbox{ , } k \rightarrow +\infty, \nonumber \\
&&u(r) = C_k F_{\ell \eta} (\Lambda_0(r)) + C_k \frac{\alpha_{k}(r)}{k^2} \mbox{ , } k \rightarrow +\infty \mbox{ , } r \rightarrow + \infty,
\label{u_large_k_Coulomb} \\
&&\mathcal{V}_0(r) = \int_{0}^{r} v_0(r')~dr'\mbox{ , } \Lambda_0(r) = \int_{0}^{r} \sqrt{k^2 - v_0(r')}~dr' \label{vm_primitive},
\end{eqnarray}
where notations are the same as in Eq.(\ref{u_large_k}), $\mathcal{V}(r)$ replaced by $\mathcal{V}_0(r)$
and $\sup_{k > k_{min}} |\alpha_k(r)| \rightarrow 0$ for $r \rightarrow +\infty$.
As analytical properties of Bessel functions are used in Sec.(\ref{compl_rel_diff_box}),
one has to express Eq.(\ref{u_large_k_Coulomb}) with Bessel functions. Using asymptotic expansions of $F_{\ell \eta} (kr)$
and derivative \cite{Abramowitz_Stegun}, one obtains:
\begin{eqnarray}
&&u(r) = C_k \hat{j}_{\ell} (kr) + C_k \frac{V_c \left[ \Psi(\ell+1) - \log(2kr) \right] - \mathcal{V}_0(r)}{2k} \hat{j}'_{\ell} (kr) 
+ O \left( C_k \frac{\log^2(k)}{k^2} \right) \mbox{ , } k \rightarrow +\infty, \nonumber \\
&&u(r) = C_k \hat{j}_{\ell} \left( \Lambda_0(r) + \frac{V_c \left[ \Psi(\ell+1) - \log(2 \Lambda_0(r)) \right]}{2k} + O(k^{-2}) \right) 
+ C_k \frac{\alpha_{k}(r)}{k^2}  \mbox{ , } k \rightarrow +\infty \mbox{ , } r \rightarrow + \infty, 
\label{u_large_k_Coulomb_Bessel}
\end{eqnarray}
where $O(k^{-2})$ is independent of $r$, $\alpha_k(r)$ is alike to that utilized in Eq.(\ref{u_large_k_Coulomb}),
$\Psi(x) = \frac{\Gamma'(x)}{\Gamma(x)}$ and $\log(k)$ terms appear in the expansion, which were absent in Eq.(\ref{u_large_k}).
They obviously do not change convergence properties of series, 
in particular Abel transformation (see App.(\ref{Abel_transformation})) still provides absolute convergence to semi-convergent series 
(see Sec.(\ref{compl_rel_diff_box}), Sec.(\ref{compl_rel_open_space}) and App.(\ref{Riemann_improper})). 
Eq.(\ref{u_large_k_Coulomb_Bessel}) is also no longer defined in $r=0$ due to $\log(2kr)$ term. 
As logarithmic divergence in Eq.(\ref{u_large_k_Coulomb_Bessel}) for $r \rightarrow 0$ occurs only in wave function normalization in Eq.(\ref{Cm_asymp}), 
it poses no problem as well because it is therein multiplied by $\hat{j}_{\ell} (kr)$. 
The only other non trivial change concerns the expansion of $k_m$
for $k_m \rightarrow +\infty$ and $R \rightarrow +\infty$ in Eq.(\ref{kn_diff_norm_large_R_large_k}), which becomes:
\begin{eqnarray}
k_m &=& \frac{(m + \frac{\ell}{2}) \pi}{R} + \frac{a_{\ell} + R \mathcal{V}_0(R) + R V_c[\Psi(\ell+1) - \log(2 k_m R)]}{2 k_m R^2} 
+ \frac{\beta_{k_m}(R)}{R k_m^2} 
\label{km_large_R_large_k}.
\end{eqnarray}
In Eq.(\ref{kn_diff_norm_large_R_large_k}), $Y_k(R)$ and $Z_k(R)$ have to be multiplied by $\log(k)^2$.

\subsection{Different potential asymptotic for $r \rightarrow 0$ and $r \rightarrow +\infty$} \label{diff_l_Vc}
From the conditions imposed to Hamiltonians in Sec.(\ref{H_pot}) and Sec.(\ref{Coulomb_div_pot_zero}), potentials bear the same centrifugal and Coulomb
coupling constants for $r \rightarrow 0$ and $r \rightarrow +\infty$:
\begin{eqnarray}
h(r) \sim \frac{\ell(\ell+1)}{r^2} + \frac{V_c}{r} \mbox{ , } r \rightarrow 0 \mbox{ or } r \rightarrow +\infty, \label{v_asymp_same}
\end{eqnarray}
where $h(r)$ is the sum of $v(r)$ and centrifugal potential.
This is obviously a nuisance for the Coulomb part, as the charge of a physical system usually changes as $r$ increases.
Fixing the centrifugal potential also prevents the consideration of potentials such as the P{\"o}schl-Teller-Ginocchio potential \cite{PTG_pot}, 
for which $\ell > 0$ in general for $r \rightarrow 0$ but $\ell = 0$ for $r \rightarrow +\infty$ in Eq.(\ref{v_asymp_same}).

This restriction is, however, not necessary in our method. To be consistent with Sec.(\ref{H_pot}), we will continue to denote
centrifugal and Coulomb coupling constants as $(\ell$, $V_c)$ for $r \rightarrow +\infty$ in Eq.(\ref{v_asymp_same}), 
but one will write them as $(\ell'$, $V_c')$ if $r \rightarrow 0$.
One can check that $(\ell'$, $V_c')$ have to be used instead of $(\ell$, $V_c)$ 
in Secs.(\ref{large_k},\ref{compl_rel_diff_box},\ref{Coulomb_div_pot_zero}),
in order to treat the first turning point occurring close to $r = 0$ correctly. 
$v(r)$ and $\mathcal{V}(r)$ must be replaced accordingly by $v_0(r)$ and $\mathcal{V}_0(r)$ defined as:
\begin{eqnarray}
v_0(r) = v(r) + \frac{\ell(\ell+1) - \ell'(\ell'+1)}{r^2} - \frac{V_c'}{r} \mbox{ , } \mathcal{V}_0(r) = \int_{0}^{r} v_0(r')~dr' \label{v_change}.
\end{eqnarray} 
$(\ell$, $V_c)$ values must be employed without change in Secs.(\ref{scat_wfs_k_zero},\ref{large_n}).
In Secs.(\ref{large_R_sec},\ref{compl_rel_open_space}), both sets of constants $(\ell$, $V_c)$ and $(\ell'$, $V_c')$ have to be used.
We will now see how to proceed in the two latter aforementioned sections.

In Sec.(\ref{large_R_sec}), $(\ell$, $V_c)$ have to be employed until Eq.(\ref{Ck_delta_large_R_large_k}), 
as one uses therein the fact that only Coulomb + centrifugal potentials remain after $R_0$. However, in order to obtain Eq.(\ref{Ck_delta_large_R_large_k}), 
one compares the wave function of Eq.(\ref{u_large_R}), defined with $(\ell$, $V_c)$, with that of Eq.(\ref{u_large_k_r}), derived with $(\ell'$, $V_c')$
constants. In Eq.(\ref{Ck_delta_large_R_large_k}), the relation between normalization constant $C_k$ and $\mathcal{N}_k$ remains the same
but $(\ell'$, $V_c')$ must be utilized for the rest of Sec.(\ref{large_R_sec}).

In Sec.(\ref{compl_rel_open_space}), $(\ell'$, $V_c')$ have to be used except for the $u_c(k_m',r)$ and $\widetilde{u}_c(k,r)$ functions,
which were defined in Sec.(\ref{scat_wfs_k_zero}) with $(\ell$, $V_c)$. All stated results remain valid.

\subsection{Complex potentials} \label{complex_pot}
One will consider the case of non-hermitian complex potentials by way of analytic continuation of Eq.(\ref{formal_conv_int}).
For this, one introduces the following complex potentials:
\begin{eqnarray}
v(r) = v_0(r) + \lambda v_1(r) \mbox{ , } w(r,r') = w_0(r,r') + \lambda w_1(r,r') 
\mbox{ , } V_c = V_{c_0} + \lambda V_{c_1} \mbox{ , } V'_c = V'_{c_0} + \lambda V'_{c_1} \label{complex_pots},
\end{eqnarray}
where $v_0(r)$, $v_1(r)$, $w_0(r,r')$, $w_1(r,r')$ are real potentials verifying Eqs.(\ref{v_asymp_inf},\ref{w_zero_cont_cond}),
$V_{c_0}$, $V'_{c_0}$ and $V_{c_1}$, $V'_{c_1}$ their associated Coulomb constants (see Eq.(\ref{v_asymp_inf}) and Sec.(\ref{diff_l_Vc}))
and $\lambda \in \C$ is the complex variable with which analytic continuation will be performed.
Angular momenta and centrifugal potential are also treated with analytic continuation:
\begin{eqnarray}
\ell = \ell_0 + \lambda \ell_1 \mbox{ , } \ell' = \ell'_0 + \lambda \ell'_1 \label{complex_l},
\end{eqnarray}
where $\ell_0 \in \R^+$, $\ell'_0 \in \R^+$ and $\ell_1 \in \R$, $\ell'_1 \in \R$. $\ell$ and $\ell'$
must respectively verify $\Re(\ell) \geq -\frac{1}{2}$ and $\Re(\ell') \geq -\frac{1}{2}$
(see Secs.(\ref{H_pot},\ref{diff_l_Vc})). 
We can assume without loss of generality that the hermitian Hamiltonian defined with $\lambda = 0$
bears no BSEC's, as this situation will be recovered with variations of $\lambda$.

For $(r,r') \in [0:+\infty[^2$ fixed, Eq.(\ref{formal_conv_int}) becomes a function of $\lambda$, denoted as $S(r,r',\lambda)$,
where tilde symbols are suppressed for convenience and $\ell'$ replaces $\ell$, in accordance with Sec.(\ref{diff_l_Vc}):
\begin{eqnarray}
S(r,r',\lambda) &=& \sum_{n=0}^{+\infty} u_n(r) u_n(r') 
         + \int_{0}^{+\infty} \left[ u(k,r) u(k,r')
         - \frac{2}{\pi} \hat{j}_{\ell'}(k r) \hat{j}_{\ell'}(k r')~dk \right], \label{S_lambda}
\end{eqnarray}
where $\lambda$-dependence is implicitly present in all wave functions by way of Eqs.(\ref{complex_pots},\ref{complex_l}).
Analytic properties of $S(r,r',\lambda)$ depends on both analytic continuation of involved wave functions with respect to $\lambda$ as well as convergence
properties of series and integral in Eq.(\ref{S_lambda}). 

As potentials can have complex values in Sec.(\ref{large_k}), Eq.(\ref{int_term_asymp})
is still available, so that the integral of Eq.(\ref{S_lambda}) converges for $k \rightarrow +\infty$, while convergence for $k \rightarrow 0^+$
of the integral poses no problem as wave functions are bounded therein, due to their normalization by way of Dirac delta normalization.
The series of Eq.(\ref{S_lambda}) has an infinite number of terms only if $\Re(V_c) < 0$. It can be handled as in Sec.(\ref{large_n}), using
wave functions $w_n(t)$ defined as $w_n(t = r |V_c|) = u_n(r)$. One verifies that the use of $(t,w(t))$ instead of $(r,u(r))$
replaces $V_c$ by $-1$ in Eq.(\ref{Schrodinger_eq}). Thus, imaginary parts of potentials and energies are put in rest terms in Sec.(\ref{large_n}), 
so that Eq.(\ref{u_large_n}) is still valid with $V_c \in \C$ verifying $\Re(V_c) < 0$. 

Existence and unicity of wave functions is provided as for the real case, using Cauchy-Lipschitz
theorem and Fredholm theory (see Sec(\ref{H_pot})), which readily defines analytic continuation in $\lambda$ of unnormalized wave functions.
Wave functions singularities occur only by way of their normalization.
Firstly, it is supposed that $u_n(r)$ functions can never become unbound or identically equal to zero with variations of $\lambda$, 
as then they cannot be normalized and hence are not analytical in $\lambda$ therein.
Secondly, analytic continuation of bound wave functions with respect to $\lambda$ implies that their normalization is effected with the square of wave functions,
and not their modulus square. Thus, if it happens that the norm of a square-integrable bound state is equal to zero,
the state cannot be normalized, i.e.~it is an exceptional point \cite{Heiss}. 
They appear with complex potentials only and are obtained by making two different bound or resonant states of a real potential
becoming degenerate by way of increase of the potential imaginary part \cite{Heiss}. This obviously occurs only for a discrete set of complex $\lambda$'s.
Another source of wave functions poles is induced by spectral singularities \cite{Samsonov}. Contrary to real potentials, complex potentials can bear
resonant states of real energy $E > 0$, for which $S^- = 0$ in Eq.(\ref{u_large_R_h_pm}). This is incompatible with the requirement that 
$S^+ S^- = \frac{1}{4}$ (see Sec.(\ref{large_R_sec})), required by Dirac delta normalization, so that they generate a singularity at $k^2 = E$. 
Contrary to exceptional points, they can be handled by deformation of the $k$-real axis in the complex plane,
where application of Cauchy theorem have them appeared as poles, similar to resonant states:
\begin{eqnarray}
S(r,r',\lambda) &=& \sum_{n=0}^{+\infty} u_n(r) u_n(r') + \sum_{m=1}^{M} u_m^{(s)}(r) u_m^{(s)}(r')
         + \int_{L^+} \left[ u(k,r) u(k,r')
         - \frac{2}{\pi} \hat{j}_{\ell'}(k r) \hat{j}_{\ell'}(k r')~dk \right], \label{S_lambda_complex_scaling}
\end{eqnarray}
where $M \in \N^*$, $u_m^{(s)}(r)$ is a properly normalized spectral singularity lying under the complex contour $L^+$, 
which can be the real axis except in the vicinity of spectral singularities,
where it consists in small half-circles in the lower complex plane encompassing $u_m^{(s)}(r)$ states.
This procedure allows to extend the domain of analyticity of $S(r,r',\lambda)$, 
and in fact generates full Berggren completeness relation \cite{Berggren} if one allows $u_m^{(s)}(r)$ to be resonant as well
(see also Ref.\cite{nikolaiev_olkhovsky} for a study of short-range complex potentials spectrum completeness properties with Berggren completeness relation).
If $u_m^{(s)}(r)$ states move to the upper part of the complex plane
for $\lambda$ close to that generating spectral singularities, 
they become bound states so that $L^{+}$ can be replaced by the real axis, recovering Eq.(\ref{S_lambda}),
but with $u_m^{(s)}(r)$ states being added to the series of Eq.(\ref{S_lambda}).
For the $\lambda$ value generating spectral singularities, the only possibility to generate a complete set of states
is to leave $L^{+}$ in the complex plane and to use Berggren completeness relation.
BSEC's can be treated in a similar manner (see Ref.\cite{Weber} for a study of BSEC's induced by non-local potentials.).

Hence, series and integrals Eq.(\ref{S_lambda}) converge normally with respect to $(r,r') \in [0:R_t]^2$, $R_t > 0$ and $\lambda$
for all $\lambda$'s preventing singularities from appearing in series and integral. Handling of singularities crossing the real axis is effected
with aforementioned contour deformation.
Thus, the analyticity domains of $S(r,r',\lambda)$ and involved complex wave functions are the same.
If $\lambda \in \R$, one is in the hermitian case, 
so that $S(r,r',\lambda)$ can be treated with Eq.(\ref{formal_conv_int}). Hence, it is identically equal to zero
in a small interval $[\lambda_{i},\lambda_{f}]$ containing $\lambda = 0$, chosen so that no bound state of energy zero occurs in this interval.
$S(r,r',\lambda) = 0$ immediately follows in its domain of analyticity. Completeness of wave functions generated by a complex potential is then procured 
as in Sec.(\ref{compl_rel_open_space}), as the set of Riccati-Bessel functions for $\ell' \in \C$, $\Re(\ell') \geq -\frac{1}{2}$,
is complete (see App.(\ref{cwf_completeness})).

\section{Conclusion}
Restricting Hamiltonians to vanish identically after a finite radius $R_0$, except for a pure Coulomb potential part, has allowed to demonstrate
the completeness of their eigenstates by way of analytical properties of Bessel and Coulomb wave functions.
It has also allowed non-local potentials to be included, as their localization before $R_0$ renders their effect negligible for $k \rightarrow +\infty$. 
The subtraction of two completeness relations of wave functions defined
with box boundary conditions, one involving the studied eigenstates and the other generated by Bessel functions, has made possible to approach
the problem with elementary methods, without recurring to Lebesgue measure theory. The fact that the obtained series converges
has a strong physical significance: completeness of eigenstates is directly related to their asymptotic
convergence to sine and cosine functions for $k \rightarrow +\infty$, whose completeness properties are those of Fourier series
and are obvious from Dirichlet kernel properties.
The singular point at $k \rightarrow 0$, due to the presence of the Coulomb potential at $r \rightarrow +\infty$, possesses the same
characteristics as that of pure Coulomb Hamiltonian, whose wave functions are analytical expressions of confluent hypergeometric functions.
Their completeness relation can thus be studied independently with complex function theory. The open radial interval case can then be procured
with the box radius sent to infinity, where all limiting processes can be handled properly due to asymptotic properties of eigenstates
for $k \rightarrow 0$ and $k \rightarrow +\infty$.

Far from the drastic approximations usually effected on Hamiltonians
to derive simple demonstrations of completeness, the set of considered Hamiltonians covers almost all physical situations assuming spherical symmetry
and pure Coulomb asymptotic, 
the difference between a potential decreasing quickly to zero or vanishing after a finite radius being nothing but academic for practical purposes.
It is hoped that this demonstration, derived with full rigor but accessible to non-specialists in operator theory, will provide better insight
to the difficult problem of Hamiltonians completeness relations of eigenstates.

\section*{Acknowledgments}
Discussions with K.~Matsuyanagi, A.~Mukhamedzanov, W.~Nazarewicz and M.~P{\l}oszajczak are gratefully acknowledged.
The author acknowledges Japan Society for the Promotion of Science for awarding The JSPS Postdoctoral Fellowship for Foreign Researchers.

\appendix

\renewcommand{\theequation}{A\arabic{equation}}
\setcounter{equation}{0}
\section{Abel transformation of semi-convergent Fourier series} \label{Abel_transformation}
The partial sums defined by the terms of Eqs.(\ref{series_term_asymp}) generating non-absolutely convergent series can be noticed to be of the form:
\begin{eqnarray}
\sum_{m=1}^{M} f(x) \frac{\sin (m x + a)}{m} \label{partial_sum_f},
\end{eqnarray}
where
\begin{eqnarray}
x = \frac{\pi (r + r')}{R} \mbox{ , } a = \ell \pi \left( \frac{r + r'}{2R} - 1 \right) \mbox{ , } 
f(x) = \frac{R(\mathcal{V}(r) + \mathcal{V}(r')) - (r + r') \mathcal{V}(R)}{2 m \pi R}, \label{xp}
\end{eqnarray}
or
\begin{eqnarray}
x = \frac{\pi (r - r')}{R} \mbox{ , } a = \ell \pi \left( \frac{r - r'}{2R} \right) \mbox{ , } 
f(x) = \frac{R(\mathcal{V}(r) - \mathcal{V}(r')) - (r - r') \mathcal{V}(R)}{2 m \pi R} \label{xm}.
\end{eqnarray}
In these definitions, $M \in \N^*$, $x \in [0:2 \pi]$ in Eq.(\ref{xp}) while $x \in [-\pi:\pi]$ in Eq.(\ref{xm}).
$f(x)$ is differentiable due to Eq.(\ref{v_primitive}) and $f(x) = 0$ if $\sin (m x + a) = 0$.
It is then convenient to use the following function $g(x)$ and partial sum $S_M(x)$:
\begin{eqnarray}
g(x) &=& \frac{f(x) e^{i (x+a)}}{1 - e^{i x}}~(x \not\in \{0,2\pi\}) \mbox{ , } g(x) = i e^{i a} f'(x)~(x \in \{0,2 \pi\}), \\
S_M(x) &=& g(x)~e^{-i x} (1 - e^{i x}) \sum_{m=1}^{M} \frac{e^{i m x}}{m} \label{SM_g},
\end{eqnarray}
where $g(x)$ is continuous on its domain of definition and Eq.(\ref{partial_sum_f}) is equal to the imaginary part of $S_M(x)$ $\forall M \in \N^*$.
One can apply Abel transformation to Eq.(\ref{SM_g}) for $M \in \N^*$:
\begin{eqnarray}
S_M(x) = g(x) \sum_{m=1}^M (1 - e^{i m x}) \left( \frac{1}{m} - \frac{1}{m+1} \right) + g(x) \frac{1 - e^{i M x}}{M+1} \label{SM_g_Abel}
\end{eqnarray}
Letting $M \rightarrow +\infty$ in Eq.(\ref{SM_g_Abel}), one obtains the value of the its limit series $S(x)$:
\begin{eqnarray}
S(x) = g(x) \sum_{m=1}^{+\infty} (1 - e^{i m x}) \left( \frac{1}{m} - \frac{1}{m+1} \right) \label{S_g_Abel}
\end{eqnarray}
$S(x)$ is normally convergent in Eq.(\ref{S_g_Abel}) with respect to $x$ and the series defined in Eq.(\ref{partial_sum_f}) 
with $M \rightarrow +\infty$ is equal to its imaginary part.
As a consequence, Eq.(\ref{series_term_asymp}) can be rewritten so as to generate a normally convergent series $\forall (r,r') \in [0:R]^2$.

\renewcommand{\theequation}{B\arabic{equation}}
\setcounter{equation}{0}
\section{Equivalent of norms of wave functions for $V_c < 0$ and $k \rightarrow 0^+$} \label{norm_equiv}
In Sec.(\ref{small_k_attractive}) and Sec.(\ref{large_n}), 
one has to devise the equivalent of the integral $\mathcal{I}$ for $r_f \rightarrow +\infty$,
representing a part of the norm of a low-energy wave function $u(r)$ in an attractive Coulomb field:
\begin{eqnarray}
u(r) \propto \lambda(r)^{-\frac{1}{2}} \left[ \sin(\Lambda(r) + \delta) + O \left( r^{-\frac{1}{2}} \right) \right] \nonumber, \\
\mathcal{I} = \int_{r_i}^{r_f} \lambda^{-1}(r) \left[ \sin^2(\Lambda(r) + \delta) + O \left( r^{-\frac{1}{2}} \right) \right]~dr \label{N_def}.
\end{eqnarray}
Eq.(\ref{N_def}) becomes after partial integration:
\begin{eqnarray}
\mathcal{I} &=& \frac{1}{2} \int_{r_i}^{r_f} \lambda^{-1}(r)~dr
-  \frac{1}{4} [\sin(2\Lambda(r) + 2\delta)~\lambda^{-2}(r)]^{r_f}_{r_i}  \nonumber \\
&+& \frac{1}{4} \int_{r_i}^{r_f} \sin(2\Lambda(r) + 2\delta) \frac{d}{dr'} \left( \lambda(r')^{-2} \right)_{r'=r}~dr
+  \int_{r_i}^{r_f} O \left( \frac{\lambda^{-1}(r)}{r^{\frac{1}{2}}} \right)~dr  \label{N_def_exp}.
\end{eqnarray}
One will show that Eq.(\ref{N_def_exp}) is equivalent to its first integral if $r_f \rightarrow +\infty$.

Let us first consider $u(r)$ to be a scattering state of linear momentum $k$ defined with box boundary conditions (see Sec.(\ref{small_k_attractive})). 
In this case, one has:
\begin{eqnarray}
r_i = R_0, r_f = R, \lambda(r) = \sqrt{k^2 + \frac{|V_c|}{r} - \frac{\ell(\ell+1)}{r^2}} \mbox{ , } \Lambda(r) = \int_{r_i}^{r} \lambda(r')~dr' \label{N_scat_def}
\end{eqnarray}
where $R_0$ is fixed and $R \rightarrow +\infty$.
Eq.(\ref{N_scat_def}) implies for $r_f \rightarrow +\infty$ and $k \in ]0:k_{\epsilon}]$, $k_{\epsilon} > 0$:
\begin{eqnarray}
\lambda^{-2}(r_f) = O(r_f) \mbox{ , } \frac{d}{dr} \left( \lambda^{-2}(r) \right)_{r=r_f} = O(1) \mbox{ , }
\int_{r_i}^{r_f} \lambda^{-1}(r)~dr = O \left( {r_f}^{\frac{3}{2}} \right) \label{lambda_scat_k_zero}.
\end{eqnarray}

If $u(r)$ is a bound state of linear momentum $k_n$ (see Sec.(\ref{large_n})), one has:
\begin{eqnarray}
r_i = r_d, r_f = r_e^{(n)}, \lambda(r) = \sqrt{\frac{|V_c|}{r} - \frac{\ell(\ell+1)}{r^2} - \kappa_n^2} \mbox{ , } \Lambda(r) = \int_{r_d}^{r} \lambda(r')~dr'
\label{N_bound_def},
\end{eqnarray}
where $r_d$ is fixed, $\kappa_n = -i k_n$ and $r_e^{(n)} = \frac{|V_c|}{2 \kappa_n^2} + O(1)$ 
is half the value of the turning point $r_t^{(n)}$ of $u(r)$, defined in Eq.(\ref{rt_def}) for $\kappa_n \rightarrow 0$. 
Owing to Eq.(\ref{N_bound_def}), one obtains, with $\kappa_n \rightarrow 0$, and thus $r_f \rightarrow +\infty$:
\begin{eqnarray}
&&\lambda(r_f) \sim \kappa_n \mbox{ , } \lambda^{-2}(r_f) \sim \kappa_n^{-2} \mbox{ , } 
\frac{d}{dr} \left( \lambda^{-2}(r) \right)_{r=r_f} = O(1), \nonumber \\
&&\int_{r_i}^{r_f} \lambda^{-1}(r)~dr = \kappa_n^{-3} 
       \int_{\kappa_n^2 r_i}^{\kappa_n^2 r_f} \left( \frac{|V_c|}{t} - 1 - \frac{\ell(\ell+1)}{t^2} \kappa_n^2 \right)^{-\frac{1}{2}}~dt
= O(\kappa_n^{-3}) \label{lambda_bound}.
\end{eqnarray}

For situations embedded in Eqs.(\ref{lambda_scat_k_zero},\ref{lambda_bound}), 
the following properties hold for $r_f \rightarrow +\infty$:
\begin{eqnarray}
&&\int_{r_i}^{r_f} \lambda^{-1}(r)~dr \rightarrow +\infty \mbox{ , } 
\lambda^{-2}(r_f) = o \left( \int_{r_i}^{r_f} \lambda^{-1}(r)~dr \right) \mbox{ , } 
\frac{d}{dr} \left( \lambda^{-2}(r) \right)_{r=r_f} = o \left( \int_{r_i}^{r_f} \lambda^{-1}(r)~dr \right), \nonumber \\ 
&&\int_{r_i}^{r_f} \frac{\lambda^{-1}(r)}{r^{\frac{1}{2}}}~dr  = o \left( \int_{r_i}^{r_f} \lambda^{-1}(r)~dr \right) \label{equiv_int_small}.
\end{eqnarray}
Standard theorems can then be used with Eqs.(\ref{N_def_exp},\ref{equiv_int_small}) to provide the result: 
\begin{eqnarray}
\int_{r_i}^{r_f} \lambda^{-1}(r) \left[ \sin^2(\Lambda(r) + \delta) + O \left( r^{-\frac{1}{2}} \right) \right]~dr 
\sim \frac{1}{2} \int_{r_i}^{r_f} \lambda^{-1}(r)~dr \mbox{ , } r_f \rightarrow +\infty  \label{norm_equivalent}.
\end{eqnarray}

\renewcommand{\theequation}{C\arabic{equation}}
\setcounter{equation}{0}
\section{Completeness relation of Coulomb wave functions in a repulsive field} \label{cwf_completeness}
In our demonstration, the completeness relation of Coulomb wave functions generated by a pure repulsive Coulomb + centrifugal potential
is required. The attractive Coulomb + centrifugal potential with $\ell \in \N$ has been treated in Ref.\cite{Mukunda}, but the repulsive case 
was not mentioned. However, due to the very similar structure of attractive and repulsive Coulomb wave functions, the demonstration of Ref.\cite{Mukunda}
just has to be mildly changed in order for the repulsive potential to be taken into account, 
as well as for $\ell \in \C$, $\Re(\ell) \geq -\frac{1}{2}$.

Following the method used in Ref.\cite{Mukunda}, one defines the integral:
\begin{eqnarray}
J(K;r,r') = \frac{2}{\pi} \int_{0}^{K} F_{\ell \eta} (kr) F_{\ell \eta} (kr')~dk \label{Coulomb_K_integral}
\end{eqnarray}
where $K > 0$, $r > 0$, $r' > 0$, $\eta$ is the Sommerfeld parameter equal to $\frac{V_c}{2k}$, $V_c \geq 0$ 
and Coulomb wave functions are properly normalized to the Dirac delta normalization. 
One demands also that $r' \geq r$, which can be clearly effected without loss of generality. 
For $K \rightarrow +\infty$, this integral will converge weakly to $\delta(r - r')$, 
as no Coulomb bound state appears in the repulsive case. Using the expression of $F_{\ell \eta}(kr)$
in terms of confluent hypergeometric functions \cite{Abramowitz_Stegun}, 
denoted as $\Phi(a,b,z)$ and $\Psi(a,b,z)$ for regular and irregular functions respectively,
Eq.(\ref{Coulomb_K_integral}) becomes \cite{Mukunda}:
\begin{eqnarray}
J(K;r,r') &=& \left( \frac{1}{2 \pi} \right) \frac{(4 r r')^{\ell + 1}}{\Gamma(2 \ell + 2)^2}
              \int_{0}^{K} \Gamma(\ell + 1 + i \eta) \Gamma(\ell + 1 - i \eta) k^{2 \ell + 2} e^{i k (r + r')} e^{-\pi \eta} \nonumber \\
          &\times& \Phi(\ell + 1 + i \eta,2 \ell + 2,-2ikr) \Phi(\ell + 1 + i \eta,2 \ell + 2,-2ikr')~dk \label{Coulomb_K_integral_expanded} \\
          &=& \left( \frac{1}{2 \pi} \right) \frac{(4 r r')^{\ell + 1}}{\Gamma(2 \ell + 2)}
              \int_{0}^{K} k^{2 \ell + 2} e^{i k r} \Phi(\ell + 1 + i \eta,2 \ell + 2,-2ikr)  \nonumber \\
          &\times& \left( e^{-i \pi (\ell+1) + i k r'} \Gamma(\ell + 1 + i \eta) \Psi(\ell + 1 + i \eta,2 \ell + 2,-2ikr')  \right. \nonumber \\
          &+& \left. e^{i \pi (\ell+1) -i k r'} \Gamma(\ell + 1 - i \eta) \Psi(\ell + 1 - i \eta,2 \ell + 2,-2ikr' e^{i \pi}) \right)~dk 
\label{Coulomb_K_integral_sym} \\
          &=& \left( \frac{1}{2 \pi} \right) \frac{(4 r r' e^{-i \pi})^{\ell + 1}}{\Gamma(2 \ell + 2)}
              \int_{-K}^{K} k^{2 \ell + 2} e^{i k (r + r')} \Phi(\ell + 1 + i \eta,2 \ell + 2,-2ikr) \nonumber \\
          &\times& \Gamma(\ell + 1 + i \eta) \Psi(\ell + 1 + i \eta,2 \ell + 2,-2ikr')~dk \label{Coulomb_K_minus_K_integral}.
\end{eqnarray}
Eq.(\ref{Coulomb_K_integral_sym}) is obtained making use of the definition of $\Psi(a,b,z)$ as a linear combination of $\Phi(a,b,z)$ functions 
for $2 \ell \not\in \N$, the case of $2 \ell \in \N$ being accounted for by a standard limiting process \cite{Abramowitz_Stegun}.
Negative $k$'s in Eq.(\ref{Coulomb_K_minus_K_integral}) are treated as $|k| e^{i \pi}$ \cite{Mukunda}, 
as one will consider in the following complex $k$'s belonging to the upper part of the complex plane only.
Note that $k^{2 \ell + 2}$ is no longer an even function of $k$ if $\ell \not\in \N$.

The integral of Eq.(\ref{Coulomb_K_minus_K_integral}) runs from $-K$ to $K$, so that one can use complex integration with a half-circle contour of radius $K$
in the upper part of the complex plane to which the real axis is added. The irregular point at $k = 0$ will be treated using a small half-circle 
of radius $k_{\epsilon}$ in the upper part of the complex plane which will be shown to vanish for $k_{\epsilon} \rightarrow 0$, 
similarly to Ref.\cite{Mukhamedzhanov}.
$e^{i k (r + r')}~\Phi(\ell + 1 + i \eta,2 \ell + 2,-2ikr)$
has a finite limit for $k \rightarrow 0$, directly determined from $\Phi(\ell + 1 + i \eta,2 \ell + 2,-2ikr)$ power series expansion \cite{Mukunda}:
\begin{eqnarray}
e^{i k (r + r')}~\Phi(\ell + 1 + i \eta,2 \ell + 2,-2ikr) &=& e^{i k (r + r')} 
\sum_{n=0}^{+\infty} \frac{(\ell + 1 + i \eta)_n}{n!~(2 \ell + 2)_n} (-2ikr)^n \nonumber \\
&=& e^{i k (r + r')}~\left[ 1 + \sum_{n=1}^{+\infty} \frac{(r V_c - 2ikr(\ell + 1)) \cdots (r V_c -2ikr(\ell + n))}{n!~(2 \ell + 2)_n} \right] \nonumber \\
&\rightarrow& \sum_{n=0}^{+\infty} \frac{(r V_c)^n}{n!~(2 \ell + 2)_n} \label{Phi_k_zero},
\end{eqnarray}
where $(a)_n = a \cdots (a+n-1)$ is the Pochhammer symbol and the series limit at $k = 0$ is effected summing the limit of all series terms at $k = 0$,
which is permitted due to normal convergence of the series in the unit circle of the $k$-complex plane.
The remaining term of the integrand of Eq.(\ref{Coulomb_K_minus_K_integral}) is treated 
expressing $\Psi(\ell + 1 + i \eta,2 \ell + 2,-2ikr')$ as a linear combination of $\Phi$'s if $2 \ell \not\in \N$
or with its logarithmic expansion if $2 \ell \in \N$ \cite{Abramowitz_Stegun,Mukunda}.
The integrand of Eq.(\ref{Coulomb_K_minus_K_integral}) is thus bounded in a complex neighborhood of $k = 0$, so that the contribution
of the mentioned small half-circle integral vanishes for $k_{\epsilon} \rightarrow 0$.

Poles of Eq.(\ref{Coulomb_K_minus_K_integral}) are generated only by
$\Gamma(\ell + 1 + i \eta)$, similarly to Ref.\cite{Mukunda}. They are immediately seen to bear negative imaginary part,
and thus no pole arises inside the used complex contour, contrary to the situation of Ref.\cite{Mukunda} where they correspond to Coulomb bound states.

For $K \rightarrow +\infty$, the integral on the half-circle of the complex contour can be calculated with expansions of
confluent hypergeometric functions for $|k| \rightarrow + \infty$ \cite{Abramowitz_Stegun,Mukunda}:
\begin{eqnarray}
\Psi(\ell + 1 + i \eta,2 \ell + 2,-2ikr') &=& (-2ikr')^{-\ell-1-i \eta} \left[ 1 + \frac{(\ell - i \eta)(\ell + 1 + i \eta)}{-2ikr'} + O(k^{-2}) \right] 
\nonumber \\
                                          &=& (-2ikr')^{-\ell-1} \left[ 1 - i \frac{V_c \log (k)}{2 k} + A k^{-1}
+ O \left( \frac{\log(k)^2}{k^2} \right) \right] \label{Psi_exp} \\
\frac{\Gamma(\ell + 1 + i \eta)}{\Gamma(2 \ell + 2)} \Phi(\ell + 1 + i \eta,2 \ell + 2,-2ikr)
\!\!\! &=& \!\!\! e^{\pm i \pi (\ell+1)} (-2ikr)^{-\ell-1}  \left[ 1 - i \frac{V_c \log (k)}{2 k} + B k^{-1}
 + O \left( \frac{\log(k)^2}{k^2} \right) \right] \nonumber \\
\!\!\! &+& \!\!\! e^{-2ikr} (-2ikr)^{-\ell-1}  \left[ 1 + i \frac{V_c \log (k)}{2 k} + C k^{-1}
 + O \left( \frac{\log(k)^2}{k^2} \right) \right] \label{Phi_exp}
\end{eqnarray}
where ``$\pm$'' is the sign of $-\Re(k)$ and $A,B$ and $C$ are complex constants independent of $k$.

Let us denote as $C(K;r,r')$ the aforementioned integral, so that $k$ therein runs on the complex contour of the half-circle,
i.e.~$k = K e^{i \theta}$ with $\theta \in [0:\pi]$. The integrand of $C(K;r,r')$ is naturally the same as in Eq.(\ref{Coulomb_K_minus_K_integral}).
Inserting Eqs.(\ref{Psi_exp},\ref{Phi_exp}) in Eq.(\ref{Coulomb_K_minus_K_integral}), the expression of $C(K;r,r')$ for $K \rightarrow +\infty$ comes forward:
\begin{eqnarray}
C(K;r,r') &=& \frac{e^{-i \pi (\ell+1)}}{2 \pi} \int_{0}^{\frac{\pi}{2}} i K e^{i K e^{i \theta} (r'+r)}
 \left[ e^{i \theta} - i \frac{V_c \log(K  e^{i \theta} )}{K} + E K^{-1} \right]~d\theta\nonumber \\
&+& \frac{e^{i \pi (\ell+1)}}{2 \pi} \int_{\frac{\pi}{2}}^{\pi} i K e^{i K e^{i \theta} (r'+r)}
 \left[ e^{i \theta} - i \frac{V_c \log(K  e^{i \theta} )}{K} + E K^{-1} \right]~d\theta \nonumber \\
&+& \frac{1}{2 \pi} \int_{0}^{\pi} i K e^{i K e^{i \theta} (r'-r)} \left[ e^{i \theta} + F K^{-1} \right]~d\theta
+ O \left( \frac{\log(K)^2}{K} \right) \label{CK_exp} \\
&=& -\frac{e^{-K(r+r')} \sin(\pi(\ell+1)) + \sin(K(r+r') - \pi(\ell+1))}{\pi(r+r')} \nonumber \\
&-& \frac{\sin(K(r'-r))}{\pi (r'-r)} + i\frac{F}{2} \delta_{r r'} + o(1) \label{CK_exp_integrated},
\end{eqnarray}
where $E$ and $F$ are constants independent of $K$ and $\theta$ and $\delta_{r r'}$ is a Kronecker delta.
Care was taken in handling powers of complex numbers when deriving Eq.(\ref{CK_exp}), using the fact that all $k$'s have positive imaginary parts.
The $r' \geq r$ condition is necessary for the exponential depending on $r'-r$ in Eq.(\ref{CK_exp}) to remain smaller than one in modulus.
Eq.(\ref{CK_exp_integrated}) implies that $C(K;r,r') \rightarrow -\delta(r-r')$ weakly when $K \rightarrow +\infty$ by way of its second and third term, 
its first term being proportional to $\delta(r+r')$, vanishing as $r+r' > 0$, while all the others are shown to go to zero
with integration by parts. Cauchy theorem then provides the result:
\begin{eqnarray}
\frac{2}{\pi} \int_{0}^{+\infty} F_{\ell \eta} (kr) F_{\ell \eta} (kr')~dk = \delta(r-r').
\end{eqnarray}
Due to the appearance of Dirichlet kernel in the demonstration, Fourier-Coulomb transform obeys the same convergence behavior as Fourier transform, i.e.
the Coulomb-Fourier expansion of a function $f(r)$ is equal to $\lim_{\delta \rightarrow 0} \frac{f(r+\delta) + f(r-\delta)}{2}$.
As $V_c$ can here be equal to zero, the completeness of spherical Riccati-Bessel functions in open radial interval has been demonstrated as well.

\renewcommand{\theequation}{D\arabic{equation}}
\setcounter{equation}{0}
\section{Convergence of series to improper integral via Riemann integral fundamental theorem} \label{Riemann_improper}
The expression of interest in Eq.(\ref{S_diff}) is the following:
\begin{eqnarray}
&& \int_{0}^{+\infty} \left[ \widetilde{u}(k,r) \widetilde{u}(k,r') - \frac{2}{\pi} \hat{j}_{\ell}(k r) \hat{j}_{\ell}(k r') \right]~dk \nonumber \\
&-& \sum_{m=1}^{+\infty} \left[ \widetilde{u}(k_m,r) \widetilde{u}(k_m,r') (k_{m} - k_{m-1})
- \frac{2}{\pi} \hat{j}_{\ell}(\kappa_m r) \hat{j}_{\ell}(\kappa_m r') (\kappa_{m} - \kappa_{m-1}) \right] \label{series_int_diff}
\end{eqnarray}
It is conveniently rewritten as:
\begin{eqnarray}
&& \int_{0}^{+\infty} [U(k) - J(k)]~dk
- \sum_{m=1}^{+\infty} \left( U(k_m) - J(k_m) \right) (k_{m} - k_{m-1}) \nonumber \\
&-& \sum_{m=1}^{+\infty} \left[ J(k_m) (k_{m} - k_{m-1}) - J(\kappa_m) (\kappa_{m} - \kappa_{m-1}) \right], \label{series_int_diff_simple}
\end{eqnarray}
where $U(k)$ and $J(k)$ are respectively $\widetilde{u}(k,r) \widetilde{u}(k,r')$ and $\frac{2}{\pi} \hat{j}_{\ell}(k r) \hat{j}_{\ell}(k r')$.
Infinite series in Eq.(\ref{series_int_diff_simple}) are convergent due to Eqs.(\ref{jl_asymp},\ref{km_kappa_m_asymp},\ref{int_term_asymp}).
Let us consider $K > 0$ and $R > 0$ and define $M_K \in \N^*$ so that $k_{M_K} \rightarrow K$ for $R \rightarrow +\infty$.
Reasoning with $\kappa_{M_K}$ as with $\kappa_{m_{\epsilon}}$ in Sec.(\ref{compl_rel_open_space}), one shows that
$\kappa_{M_K} \rightarrow K$ for $R \rightarrow +\infty$. 

We will focus first on the rests of the series of Eq.(\ref{series_int_diff_simple}) with respect to $M_K$:
\begin{eqnarray}
&&R_U = \sum_{m = M_K}^{+\infty} \left[ U(k_m) - J(k_m) \right] (k_{m} - k_{m-1}), \label{R_U} \\
&&R_J = \sum_{m = M_K}^{+\infty} \left[ J(k_m) (k_{m} - k_{m-1}) - J(\kappa_m) (\kappa_{m} - \kappa_{m-1}) \right] \label{R_J}.
\end{eqnarray}
It is convenient to transform the series of Eqs.(\ref{R_U},\ref{R_J}) by way of an Abel transformation (see App.(\ref{Abel_transformation}))
so that they bear absolute convergence.
We first study Eq.(\ref{R_U}).
Using Eq.(\ref{int_term_asymp}), the semi-convergent part of Eq.(\ref{R_U}) can be written as the imaginary part of the sum of two terms of the following form:
\begin{eqnarray}
F_U(x) = \sum_{m=M_K}^{+\infty} A_m(x) \left( \frac{1}{k_{m}} - \frac{1}{k_{m+1}} \right) \mbox{ , }
A_m(x) = \sum_{n=M_K}^{m} f(x) e^{i k_n x} (k_{n} - k_{n-1}), \label{R_Am_U_Abel} 
\end{eqnarray}
where $x$ and $f(x)$ are equal respectively to $r \pm r'$ and $\mp [\mathcal{V}(r) \pm \mathcal{V}(r')]/2 \pi$,
with the last term multiplied by $e^{-i \ell \pi}$ if $x = r + r'$.
$A_m(x)$ in Eq.(\ref{R_Am_U_Abel}) has an simple asymptotic expansion for $R \rightarrow +\infty$, 
deduced from Eq.(\ref{kn_diff_norm_large_R_large_k}):
\begin{eqnarray}
A_m(x) &=& \sum_{n=M_K}^{m} f(x) e^{i k_n x} \left[ \frac{\pi}{R} + \frac{Y_{k_n(R)}}{k_n^2 R} \right] \nonumber  \\
       &=& \frac{\pi}{R} e^{\frac{i \ell \pi}{2R}x}
           \sum_{n=M_K}^{m} f(x) e^{i \frac{n \pi}{R} x} \left( 1 + i \frac{a_{\ell} x}{2 R n \pi} + i \frac{\mathcal{V}(R) x}{2 n \pi} \right) + \alpha_{k_m}(R) 
           \nonumber \\
       &=& \frac{f(x)}{i x} \left( e^{i k_m x} - e^{i K x} \right) + \alpha_{k_m}(R), \label{R_Am_U_Abel_asymp} 
\end{eqnarray}
where $\sup_{k > K} |\alpha_k(R)| \rightarrow 0$ for $R \rightarrow +\infty$ and the fact that $Y_{k_n}(R) = O((k_n - k_{n-1})\log(R))$ for $R \rightarrow +\infty$
has been used.
The part of the sum in Eq.(\ref{R_Am_U_Abel_asymp}) depending on $n^{-1}$ can be handled with Abel transformation as well.
From Eq.(\ref{R_Am_U_Abel_asymp}), 
$A_m(x)$ is bounded for $k_m \rightarrow +\infty$ and $x \in \R$, as $\frac{f(x)}{x}$ is finite for $x \rightarrow 0$.
As a consequence, Abel transformation performed on Eq.(\ref{R_U}) provides the inequalities:
\begin{eqnarray}
|R_U| &\leq& A_0 \sum_{m=M_K}^{+\infty} \left[ \left( \frac{1}{k_{m}} - \frac{1}{k_{m+1}} \right) + O \left( \frac{k_{m} - k_{m-1}}{k_m^2} \right) \right]
          \leq \sum_{m=M_K}^{+\infty} \left[ O \left( \frac{1}{k_{m}} - \frac{1}{k_{m+1}} \right) + O \left( \frac{1}{k_{m-1}} - \frac{1}{k_{m}} \right) \right]
\nonumber \\
          &\leq& \frac{A}{K}, \label{R_U_major} 
\end{eqnarray}
where the property that $k_{m} > k_{m-1}$ $\forall m \in \N^*$ has been used and $A_0,A$ are positive constants independent of $K$ and $R$.
Using Eqs.(\ref{jl_asymp}) and (\ref{kn_diff_norm_large_R_large_k}), one can rewrite Eq.(\ref{R_J}) as:
\begin{eqnarray}
R_J &=&  \frac{\mathcal{V}(R)}{2 \pi R} \sum_{m=M_K}^{+\infty} \left[ \frac{(r+r') \sin(\kappa_m(r+r') - \ell \pi)
    - (r-r') \sin(\kappa_m(r-r'))}{\kappa_m} + O(\kappa_m^{-2}) \right]  (\kappa_{m} - \kappa_{m-1}) \nonumber \\
    &+& \sum_{m=M_K}^{+\infty} \frac{\alpha_{\kappa_m}(R)}{R \kappa_m^{2}} 
\label{R_J_asymp},
\end{eqnarray}
where $\sup_{\kappa > K} |\alpha_\kappa(R)| \rightarrow 0$ for $R \rightarrow +\infty$.
Abel transformation of the semi-convergent term of $R_J$ in Eq.(\ref{R_J_asymp}) can be accounted for in the same manner as for $R_U$.
Consequently, $|R_U| + |R_J|$ can be majored by $\frac{A}{K}$, where $A$ is a positive constant independent of $K$ and $R$.
We have demonstrated that Eq.(\ref{series_int_diff_simple}) can be majored by the positive constant $P$:
\begin{eqnarray}
P &=& \left| \int_{0}^{K} [U(k) - J(k)]~dk - 
      \sum_{m=1}^{M_K} \left[ U(k_m) (k_{m} - k_{m-1}) - J(\kappa_m) (\kappa_{m} - \kappa_{m-1}) \right] \right| 
      \nonumber \\
  &+& \left| \int_{K}^{\infty} [U(k) - J(k)]~dk \right| + \frac{A}{K}. \label{series_int_diff_majoration}
\end{eqnarray}
Let us take $\epsilon > 0$. The two last terms of Eq.(\ref{series_int_diff_majoration}) can be made smaller than $\epsilon$ for $K$ large enough,
independently of $R$. The remaining term of Eq.(\ref{series_int_diff_majoration}) is the difference between a Riemann sum and its integral limit
on the finite interval $[0:K]$ for $R \rightarrow +\infty$. Thus, it can made smaller than $\epsilon$ for $R$ large enough.
Eq.(\ref{series_int_diff}) has thus been shown to go to zero when $R \rightarrow +\infty$.

\renewcommand{\theequation}{E\arabic{equation}}
\setcounter{equation}{0}
\section{Limit of series $S^{\epsilon +}_{l}(r,r')$ for $R \rightarrow +\infty$} \label{norm_diff_series}
One considers the series $S^{\epsilon +}_{l}(r,r')$ defined in Sec.(\ref{compl_rel_open_space}):
\begin{eqnarray}
S^{\epsilon +}_{l}(r,r') &=& 
\sum_{m=m_{\epsilon}}^{+\infty} \left[ \frac{}{} \left( \mathcal{N}_{k_m}^2 - 1 \right) \widetilde{u}(k_m,r) \widetilde{u}(k_m,r') (k_{m} - k_{m-1}) 
\right. \nonumber \\
&-&  \left. \left( B_{\kappa_m}^2 - \frac{2}{\pi} \right) 
\hat{j}_{\ell}(\kappa_m r) \hat{j}_{\ell}(\kappa_m r') (\kappa_{m} - \kappa_{m-1}) \right], \label{Sl_eps_plus_value}
\end{eqnarray}
where $\widetilde{u}(k_m,r)$ is defined by way of Eq.(\ref{formal_conv_int}) in Sec.(\ref{compl_rel_open_space}), 
$\hat{j}_{\ell}(\kappa_m r)$ was introduced with Eq.(\ref{Bessel_comp_rel}) in Sec.(\ref{compl_rel_diff_box}),
$m_{\epsilon} \in \N^*$ is defined so that $k_{m_{\epsilon}} \rightarrow k_{\epsilon}$ 
and $\kappa_{m_{\epsilon}} \rightarrow k_{\epsilon}$ for $R \rightarrow +\infty$, $k_{\epsilon}$ being a positive linear momentum 
(see Sec.(\ref{compl_rel_open_space})) and $B_{\kappa_m}$, $\mathcal{N}_{k_m}$ are respectively defined in Eq.(\ref{Bessel_km_norm}) and Eq.(\ref{u_large_R_h_pm}).

Due to boundedness of scattering wave functions and absolute convergence of series 
(see Eq.(\ref{Cm_Bm_two_pi})), $|S^{\epsilon +}_{l}(r,r')|$ verifies the inequality:
\begin{eqnarray}
|S^{\epsilon +}_{l}(r,r')| \leq M_u \sum_{m=m_{\epsilon}}^{+\infty} |\mathcal{N}_{k_m}^2 - 1| (k_{m} - k_{m-1})
+ M_j \sum_{m=m_{\epsilon}}^{+\infty} \left| B_{\kappa_m}^2 - \frac{2}{\pi} \right| (\kappa_{m} - \kappa_{m-1}), \label{Sl_eps_plus_ineq}
\end{eqnarray}
where $M_u$ and $M_j$ are two positive constants independent of the box radius $R$.
From Eqs.(\ref{kn_diff_norm_large_R_large_k},\ref{Bessel_large_R_large_k}), $\mathcal{N}_{k_m}$ and $B_{\kappa_m}$ verify:
\begin{eqnarray}
|\mathcal{N}_{k_m}^2 - 1| \leq \frac{M_{\mathcal{N}}(R)}{k_{m}^2} \mbox{ , } 
\left| B_{\kappa_m}^2 - \frac{2}{\pi} \right| \leq \frac{M_B(R)}{\kappa_{m}^2}, \label{NB_ineq}
\end{eqnarray}
where $k_{m} > k_{\epsilon}$, $\kappa_{m} > k_{\epsilon}$,
and $M_{\mathcal{N}}(R)$, $M_B(R)$ are two positive constants independent of $k_{m}$ and $\kappa_{m}$ and vanishing for $R \rightarrow +\infty$.
Eq.(\ref{Sl_eps_plus_ineq}) implies for $S^{\epsilon +}_{l}(r,r')$ and $R$ sufficiently large:
\begin{eqnarray}
|S^{\epsilon +}_{l}(r,r')|
&\leq& M_u~M_{\mathcal{N}}(R) \sum_{m=m_{\epsilon}}^{+\infty} \frac{k_{m} - k_{m-1}}{k_{m}^2}
+ M_j~M_B(R) \sum_{m=m_{\epsilon}}^{+\infty} \frac{\kappa_{m} - \kappa_{m-1}}{\kappa_{m}^2} \nonumber \\ 
&\leq& M_u~M_{\mathcal{N}}(R) \sum_{m=m_{\epsilon}}^{+\infty} \frac{k_{m} - k_{m-1}}{k_m k_{m-1}}
+ M_j~M_B(R) \sum_{m=m_{\epsilon}}^{+\infty} \frac{\kappa_{m} - \kappa_{m-1}}{\kappa_m \kappa_{m-1}} \nonumber \\
&\leq& M_u~M_{\mathcal{N}}(R) \sum_{m=m_{\epsilon}}^{+\infty} \left( \frac{1}{k_{m-1}} - \frac{1}{k_{m}} \right)
+ M_j~M_B(R) \sum_{m=m_{\epsilon}}^{+\infty} \left( \frac{1}{\kappa_{m-1}} - \frac{1}{\kappa_{m}} \right), \nonumber \\
&\leq& 2 \frac{M_u~M_{\mathcal{N}}(R) + M_j~M_B(R)}{k_{\epsilon}},
\end{eqnarray}
where one has used the properties $k_{m} > k_{m-1}$ and $\kappa_{m} > \kappa_{m-1}$.
Hence, $S^{\epsilon +}_{l}(r,r') \rightarrow 0$ for $R \rightarrow +\infty$ $\forall~k_{\epsilon} > 0$.

\end{document}